\documentclass{article}
\usepackage{subfigure}

\usepackage{graphicx}
\usepackage{fullpage}
\usepackage{pdfpages}

\usepackage{xspace}
\usepackage{xr}
\usepackage{url}
\usepackage{xcolor}
\usepackage{caption}
\usepackage{comment}

\makeatletter
\newcommand*{\addFileDependency}[1]{%
  \typeout{(#1)}
  \@addtofilelist{#1}
  \IfFileExists{#1}{}{\typeout{No file #1.}}
}
\makeatother

\newcommand*{\myexternaldocument}[1]{%
    \externaldocument{#1}%
    \addFileDependency{#1.tex}%
    \addFileDependency{#1.aux}%
}

\myexternaldocument{supplementary}

\usepackage{xspace}
\newcommand{\ignore}[1]{}
\usepackage{color}
\newcommand{\vir}[1]{``#1''}
\newcommand{\kmer}{$k$-mer\xspace}

\newcommand{\SM}{Supplementary Material\xspace}

\newcommand{\ChiSquare}{$\chi^2$\xspace}
\newcommand{\Dd}{$D_2$\xspace}
\newcommand{\Dds}{$D^S_2$\xspace}
\newcommand{\Ddstar}{$D^*_2$\xspace}
\newcommand{\Ddz}{$D_2z$\xspace}

\title{The Power of Word-Frequency Based Alignment-Free Functions: a Comprehensive Large-scale Experimental Analysis - Version 3}

\author{
Giuseppe Cattaneo\thanks{Dipartimento di Informatica, Universit\`{a} di Salerno, Fisciano (SA), 84084, Italy} \hspace{0.05mm} \thanks{To whom correspondence should be addressed.} \and
Umberto Ferraro Petrillo\thanks{Dipartimento di Scienze Statistiche, Universit\`{a} di Roma - La Sapienza, Rome, 00185, Italy} \hspace{0.05mm} \footnotemark[1] \and 
Raffaele Giancarlo\thanks{Dipartimento  di Matematica ed Informatica, Universit\`{a} di Palermo, Palermo, 90133, Italy} \hspace{0.05mm}
\and 
Francesco Palini \footnotemark[1] 
\and 
Chiara Romualdi\thanks{Dipartimento di Biologia, Universit\`{a} di Padova, Padova, 35131, Italy} \hspace{0.05mm} 
}

\date{}

\begin{document}

\maketitle

\begin{abstract}

\textbf{Motivation:   } Alignment-free (AF) distance/similarity functions are a key tool for sequence analysis. Experimental studies on real datasets abound and, to some extent, there are also studies regarding their control of false positive rate (Type I error). However, assessment of their power, i.e., their ability to identify true similarity,  has been limited to some members of the \Dd family  by  experimental studies   on short sequences, not adequate for current  applications, {\color{black}{where sequence lengths may vary considerably.}} Such a State of the Art is methodologically problematic, since information regarding  a key feature such as power is either missing or limited.    \\
\textbf{Results:} {\color{black} By concentrating on a representative set of word-frequency based AF functions}, we perform the first coherent and uniform evaluation of the power, involving also  Type I error for completeness. Two Alternative models of important genomic features (CIS Regulatory Modules  and Horizontal Gene Transfer), a wide range of sequence lengths from a few thousand to millions, and different values of $k$ have been used. As a result, we provide a characterization of those AF functions that is novel and informative. Indeed, we identify  weak and strong  points of each function considered, which  may be used as a guide to choose one for analysis tasks. Remarkably, of the fifteen  functions that we have considered, {\color{black}{only four stand out, with small differences between small and short sequence length scenarios}}. Finally, in order to encourage the use of our methodology for validation of future AF functions, the Big Data platform supporting it is public. 

\end{abstract}

\section{Introduction}

Alignment-free distance/similarity functions (AF functions, for short) provide an alternative approach to traditional sequence alignment methods, e.g.,  \cite{altschul1990basic,smith1981identification}, for determining how distant/close two sequences are. 
Their  advantages/disadvantages with respect to alignment methods are well presented in  \cite{Zielezinski2019}.

Although AF functions have a long and established history \cite{Vinga03} and their use has been widely investigated for sequence analysis in genomics \cite{utro2019,Zielezinski2019}, metagenomics \cite{Benoit16}, and epigenomics \cite{Giancarlo2015, Giancarlo18},  only recently a comprehensive experimental study on benchmark and real datasets  has appeared in the Literature \cite{Zielezinski2019}. It has been followed by another study (as an application of the Big Data platform FADE \cite{FADE}), in which Type I error control i.e, the  ability to control the number of  false positives, has been evaluated for some prominent AF functions on the same real datasets.

However, for a full assessment of the \vir{value} of an AF function, its power, i.e, its  ability to identify true positives, must also be estimated. Unfortunately, those studies are scarce and confined to the \Dd family  or its variants. \textcolor{black}{This is an important pitfall of the methodology regarding AF functions as when the AF measure is not able to recognize a true similarity (because of poor statistical power), it will inevitably lead to a biased  result in applications, with the user not even being aware of where the problem may lie. Therefore, although our study follows in terms of time more \vir{application oriented benchmarkings}, methodologically it is a much needed prerequisite to them. }

We recall that Power estimation has been investigated  theoretically in \cite{wan2010alignment} and experimentally,  exclusively \cite{HUANG19,2009reinertalignment} or mostly \cite{Liu2011},   on synthetic datasets.  Indeed, as opposed to Type I error control studies, power studies are less amenable to \textcolor{black}{being} performed on real datasets, where they  may provide inaccurate results, as reported in \cite{Liu2011}. Therefore,  in agreement with the indications of that study,  we resort to use simulated datasets. It is to be highlighted that there is also a subtle connection between power and Type I error control. Indeed, a high power may be determined by a poor Type I error control. As a consequence, although Type I error control studies for AF functions are now available on benchmark real datasets \cite{FADE}, here we repeat them on synthetic datasets for completeness. 
We also adhere to the ground-breaking methodology proposed in \cite{2009reinertalignment}. In particular, we use two models of \vir{biologically relevant } similarity, one resulting from the process of horizontal gene transfer and the other resulting from the process of acquiring common CIS regulatory elements \cite{HUANG19,Liu2011,2009reinertalignment}. The first is referred  to as Alternative  model Pattern Transfer and the second as Motif Replace. 

{\color{black}{We consider two scenarios: short sequences (length up to the thousands, recalling the similarity of gene sequencing) and long sequences (length up to the millions, recalling the similarity of genome contigs). This gives rise to a range of experiments}} much broader than those of  \cite{Liu2011,2009reinertalignment}.  {\color{black}{ The corresponding  analysis provides the first extensive study of the power of AF functions,} } by concentrating on {\color{black}{word-frequency}} based ones \cite{luczak2017survey}. They use {\em \kmer statistics}, and  our choice is due to their simplicity, effectiveness  and widespread use, as documented by a recent benchmarking study \cite{Zielezinski2019}.
We have chosen the best performing {\color{black}{word-frequency based}} AF functions according to the mentioned benchmarking, representatives of all types of AF functions described in \cite{luczak2017survey}. \textcolor{black}{We exclude from our study those AF functions that are based only on the presence/absence  of words, e.g. Jaccard and its approximation Mash \cite{Ondov2016}, (see Supplementary File in \cite{Lu17} for definitions) since the experimental plan  by Reinert et al.,  and used  here,   turns out not to be fully adequate to study these functions as for the values of $k$ recommended in  the mentioned study, their dictionary includes all possible k-mers. 
%Such a finding gives rise to an interesting open problem. 
}
Our results are as follows.

\begin{itemize}
    
\item {\color{black}{Word-frequency based}} AF functions assure a good Type I error control. {\color{black}{That is, in a \vir{neutral environment} where we compare randomly generated pairs of sequences, the functions are difficult to fool, i.e. they found significant similarity pairs close to the expected proportion given by  the intrinsic uncertainty of the statistical test }} As shown by \cite{FADE},  when using real datasets,  Type I error control is more heterogeneous across AF functions. 

\item Our power studies provide a novel classification of {\color{black}{word-frequency}} based AF functions much more informative than the standard taxonomic one, i.e, the one provided in \cite{luczak2017survey} (see also \cite{Bernard16}). 
Briefly, we identify a handful of functions that stand out in terms of performance and  we advance the knowledge regarding the power of the   prominent and much studied \Dd family in relation to the two Alternative  models. 

\item Our results indicate that the heuristic law  commonly used for the selection of $k$, i.e., as the logarithm of sequence lengths, is generally not appropriate. In fact, for Type I error control, the choice of $k$ seems to be marginal for a good performance, while it has a complex relation with Alternative model, sequence length and function. {\color{black}{}Differently from other studies, our work provides a guideline regarding the most appropriate choice of $k$ which is very much dependent on the function and its parameters}. 

\item In order to encourage the use of our methodology for future benchmarking,  we provide a new computational framework based on a succinct data structures Big data Platform, namely FADE \cite{FADE},  which guarantees the reproducibility of the entire methodology and can be further extended with user-provided  functions.

\end{itemize}

\section{Methods}

\subsection{Distance Functions }

The {\color{black}{word-frequency} based}  AF functions chosen for this study are summarized in Section 1 of the \SM, together with their definitions. They all depend on the choice of $k$, i.e,. the $k$-mer length used for the statistics. The heuristic for the selection of $k$ is to choose as $k$  the logarithm of the sequence length (see formula 56 in \cite{luczak2017survey}).  Here we select $k$ as a set of values upper bounded by the logarithm of the maximum sequence length. Then, for the evaluation of AF function performance,  we proceed combinatorially, i.e. the assessment is made for each combination of the chosen values  of $n$ and $k$.  
        
\subsection{The Methodology for the Experimental Study of Control of Type I Error and Power}
\label{subsec:T1}
% descrizione del null model (uniform) e dei due alternative models (pattern transfert and motif replace)

A consolidated approach for the evaluation of the performance of a  statistic, in our case an AF function,  concerns the efficacy in discriminating random from real effects. {\color{black}{If we define a positive test the identification of  significant similarity,}} the corresponding quantification is generally obtained by evaluating the control of Type I error {\color{black}{(i.e. the control of the number of false positives)}} and the  power of the test statistic {\color{black}{(i.e. the identification of the true positives)}}. This  hinges on two main ingredients: Generative Models {\color{black}{to produce pairs of sequences with a given level of  similarity}}  and statistical test. 
The work by Reinert et al. \cite{2009reinertalignment} regarding the \Dd statistics is a clear instance  of such a  methodology that we closely follow, with the addition of  providing public software and benchmark datasets. 

Traditionally, the AF functions included in this study are grouped by mathematical families (see supplementary Material and \cite{Bernard16,luczak2017survey}). We anticipate that, in what follows,  functions of the same family are listed contiguously in Tables and Figures,  with no separation among families.

%\rg{menzionare il simulatore-ovvero il software}
\subsubsection{Generative Models for Sequences}\label{models}

\paragraph{Null Model.}\label{null}
It is intended to formalize the generation of pairs  of sequences  \vir{similar by chance}.  
Given the alphabet $\Phi=\{A,C,G,T\}$ with uniform probability distribution  and two integers  $n$ and $m$, $m$ pairs of sequences,   each of length $n$, are generated using random samplings from a multinomial distribution $Multinom(p,n)$. In what follows, we refer to this model simply as $NM$. {\color{black}{Moreover, for the convenience of the reader, we recall that a multinomial is a generalization of the binomial distribution when the number of events is greater than two.  }}

\paragraph{Alternative Model: Pattern Transfer. }\label{Alternative1}
This model, introduced in  \cite{2009reinertalignment} and with variants proposed in  \cite{HUANG19,kai2013}, is intended to capture the process of acquiring similarity between two biological sequences via horizontal gene transfer.  Our model is essentially the same as the original one.
{\color{black}{Informally}}, we choose at random a number of positions in one  of the two sequences. For each position, we copy the subsequence starting at that index (simply referred to as motif) in the chosen sequence, in the same position of the other sequence. The random position selection process is governed  by a parameter $\gamma\in [0,1]$. Higher the value of $\gamma$, more positions are selected.  Formal details are available in Section 2.1 of the Supplementary Material.

\paragraph{Alternative Model: Motif Replace.}
This model has been  also introduced in  \cite{2009reinertalignment}. Although no motivation was given, it intuitively represents the notion of similarity between sequences that share many Transcription  Factor binding sites, and as pointed out in 
\cite{HUANG19}, it models the acquisition of common CIS Regulatory Modules.  Technically, It is similar to $PT$, except for the selection and replacement of the motif. This latter is  selected uniformly and at random from a set of motifs. As for replacement, rather than a pattern transfer, both sequences acquire a copy of the selected motif starting at the same position.  The random position selection process is governed  by a parameter $\gamma$, as in the previous model. Formal details are available in Section 2.2 of the Supplementary Material. 

%In order to better model the biology underlying process, we modify the model by using a set of motifs to implant rather than one.  

\subsubsection{Alternative vs Null Models: the Ability to Capture Similarity Trends } \label{sec:KS}

To fix ideas, we present the case of  similarity functions, since the case of distances is analogous. Independently of whether or not a similarity function  $D$ is based on alignments, ideally, it has to be able to 
separate, in terms of its value, \vir{truly} similar sequences from those that are not. 
%{\color{black}{In practice, we expect to observe a trend in the distribution of its values that amplifies the separation as the similarity between pairs of sequences increases.}}
%In practice, it  is able to capture such a separation in terms of a trend, rather than exactly,  in a dataset of pairs of sequences. That is,  a similarity trend in the data  is  captured in terms of a separation one. 

The \vir{comparison} of  Alternative  models, {\color{black}{ that generate truly similar sequence pairs,}} with respect to Null ones, {\color{black}{ that generate random sequence pairs,}} allows for the quantification of how well a function captures these separations. 
Briefly and informally, for each AF measure, $k$ and $n$ we calculate the differences between the average AF value distribution of the NM and AMs. Higher the difference, higher the separation.  

The formal methodology we propose for such a quantification is reported in Section 3 of the Supplementary Material. 

%follows using, to fix ideas, $NM$ and $PT$ with parameter $\gamma$. 

%Formal details are available in Section \ref{sm:sec:KS} of the Supplementary Material. 

\ignore{
\begin{itemize}
    \item [(1)] Generate a set $S$ of $m$ pairs of sequences, each of length $n$,  via $NM$. Using $S$, generate a set $S_{\gamma}$ of $m$ pairs of sequences, each of length $n$,  via $PT$ with parameter $\gamma$. 
      \item [(2)]  Compute $D$ for each pair of sequences in $S$ and $S_{\gamma}$, respectively,  and in agreement with the function parameters, e.g., $k$.  
   
    \item [(3)] $AF$ values computed in (2) are reported as boxplots and also as  delta values, i.e differences between $AM$ and $NM$ average values, those latter in the form of   heatmaps.
\end{itemize}

It is to be pointed out that $S_{\gamma}$ can be computed with different values of $\gamma$. It is expected that  a good similarity function would return large delta and non-overlapping boxplots, with the ones of $S_{\gamma}$ \vir{higher} than those of $S$. 
}

\subsubsection{ {\color{black}{How to Estimate}}  Type I Error Control and Statistical Power}\label{sec:TIC}

Given an AF function, all the statistical tests we perform  in this study, and detailed next, are based on the same inferential hypotheses: $H_0$, the two sequences are similar by chance (Null hypothesis) and, $H_1$, the two sequences are more similar than they would be by chance (Alternative hypothesis). \textcolor{black}{The statistical test allows us to decide which hypothesis is more likely to be true, according to our data. Hereafter,  we define as positive a test for which the null hypothesis is likely to be false. Two different errors are associated with this decision, the Type I error, i.e. the probability of identifying  a positive event when it is actually false, and the Type II error, i.e. the probability of not identifying a positive event when it is actually true. }

\textcolor{black}{There is an inevitable trade-off between the two errors, if Type I error increases, Type II decreases, and viceversa. The general approach is to fix Type I to a defined small value (called nominal value, usually $\alpha=0.05$) and minimize Type II accordingly. Thus,  in the case of absence of true positives,  it is expected that a statistical test would identify 5\% of positives by chance, any deviation from 5\% results in a  too conservative (lower than 5\%) or too liberal (higher than 5\%) test.}

\textcolor{black}{One minus the Type II error is \emph{formally} the statistical power of a test, i.e. the probability to identify true positive events. Higher the power, better the test. }

\textcolor{black}{The quantification of both Type I error  (in case of absence of true positives) and the statistical power (in case of presence of true positives) is used to measure the performance of a statistical test.} 

\textcolor{black}{Given our generative models, $NM$ is used to define the null distribution (random sequences), while $PT$ and $MR$ are used to generate pairs  of truly similar sequences (the true positive pairs). 
Given a set of sequences generated by $NM$ and $AMs$, for each  AF function we calculate the number of significantly similar sequence pairs, that is the false positives and the true positives respectively. In Section 4 of the Supplementary Material, the test procedure is highlighted.}

\subsection{Benchmarking Software and Datasets}

The following tools, used  for this research,  have been made publicly available.

% except for the part related to the evaluation of AF measures. 

\begin{itemize}
    \item {\bf Datasets Generation. } We provide  a Spark-based distributed tool for the automatic generation of collection of genomic sequences according to the types of Generative Models described in Section \ref{models}. The tool accepts as parameters the same ones  as the models.  Output sequences are encoded as standard FASTA files.

    \item {\bf AF Function  Evaluation.} An AF function  between pair of sequences is  evaluated using the FADE software (see \cite{FADE}). It is a Spark-based distributed framework for AF analysis over large collections of genomic sequences, coming with the implementation of several popular AF functions. It rests on a multi-criteria succinct representation  of $k$-mer dictionaries, well suited for effective storage and load balancing in a distributed setting.  
    
    \item {\bf AF Functions  Analysis.} 
    We provide a Java-based tool for estimating the Type I error rate and the power of the test statistic over a set of input AF functions, according to the methodology described in Section \ref{sec:TIC}. This is done by comparatively analyzing the AF functions,  evaluated over sequences generated according to the Null model, against the corresponding sequences generated according to the PT and MR Alternative models.  Once available, these results are summarized and visualized using a proper graphical representation, by means of a collection of R scripts.
\end{itemize}

\section{Results}

\subsection{Type I Error Control}\label{sec:resT1}
For each AF function, we have estimated the Type I error rate, as outlined  in Section \ref{sec:TIC}. 
The relevant parameters that specify the set of experiments carried out are as follows. \textcolor{black}{For both scenarios (long and short sequences),  we use  $50$ different  lengths $n$ and $m=1000$ pairs of sequences. For long sequences, $n$ ranges from two hundred thousands to  ten millions, with a step of two hundred thousands,  while for short sequences $n$ ranges  from one thousand to fifty thousands with a step of one thousand. } The false positive rate has been estimated for $k=4,6,8,10$.  For  $\alpha=0.05$ \textcolor{black}{and long sequences}, the percentage of false positives are reported cumulatively by length with boxplots in Figure \ref{fig:PanelT1}.  \textcolor{black}{ For $\alpha=0.01,0.10$, the results are reported in Figure 1 of the Supplementary Material. The results for short sequences are reported in Figure 2 of the Supplementary Material.} \textcolor{black}{For each scenario, we can draw the following conclusions}. 
With the exception of  Chebyshev, each AF function  performs  fairly well,  since the values of the percentage of false positives is close to the nominal level $\alpha$.

\begin{figure}[htp] % not h only
\centering
    \includegraphics[width=0.48\textwidth]{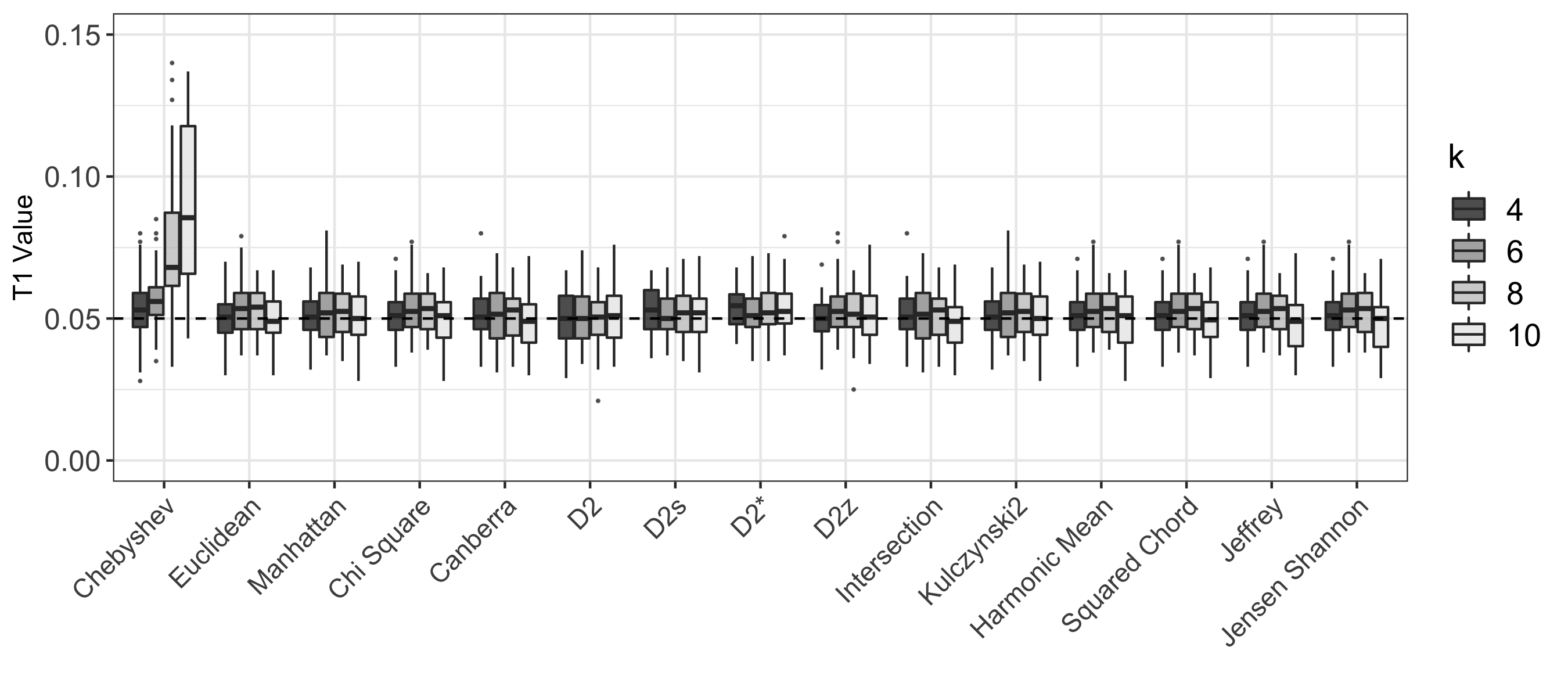}%
    \caption{Results of the Type I Error Control experiments,  \textcolor{black}{for the long sequences scenario}, nominal level $\alpha=0.05$. For each AF function (x axis) and each value of $k$, the percentage of false positives over all lengths is represented by boxplots. The dotted black line corresponds to the nominal level.}
    \label{fig:PanelT1}
\end{figure}

\subsection{Ability to Capture Similarity Trends}
\label{sec:capture}
Given a length $n=5\ 000\ 000$, and a similarity/distance function, we generate 1000 sequence pairs and process them as reported in Section \ref{sec:KS} for $NM$, $PT$, and $MR$ with $\gamma=0.01,0.05,0.1$, respectively.

\begin{figure}[htp] % not h only 
\centering
    \includegraphics[width=0.48\textwidth] {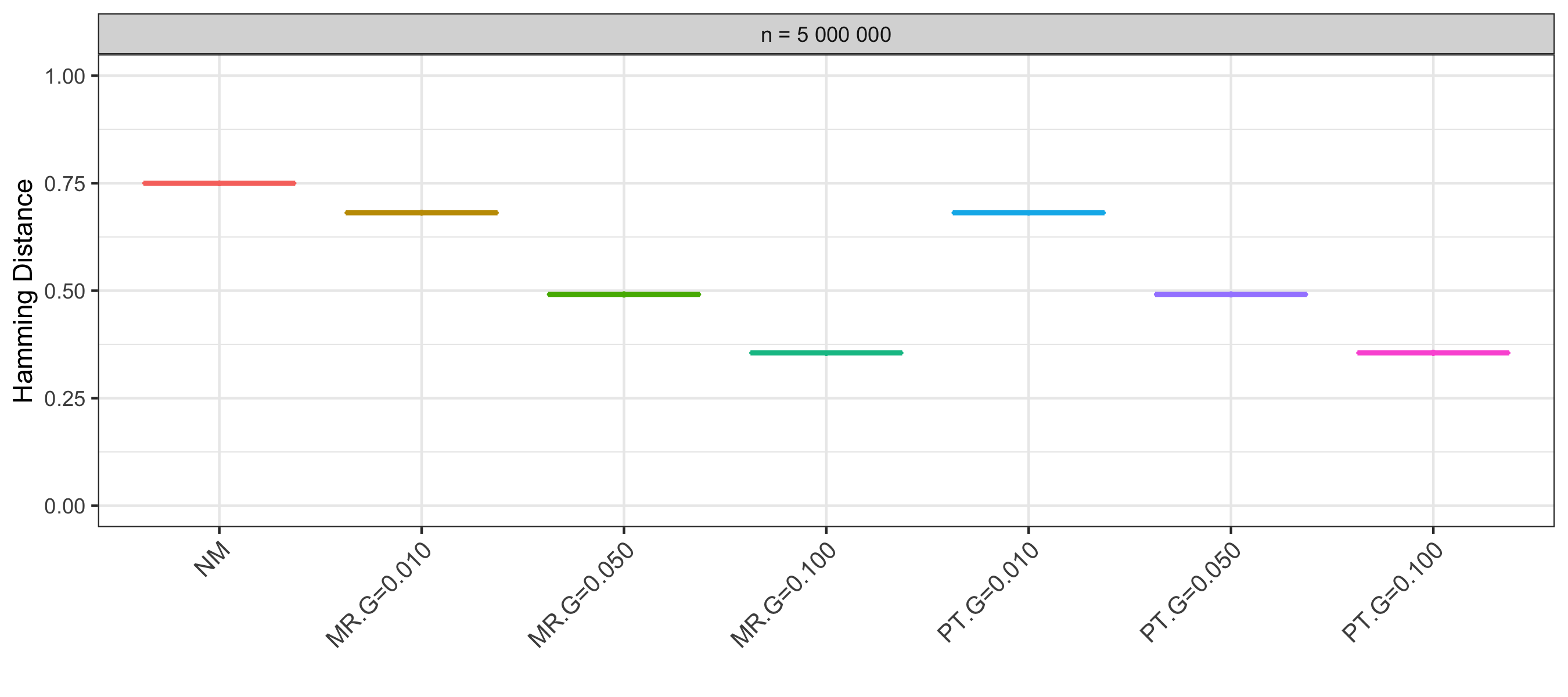}%
    \captionof{figure}{Separation. Boxplots of  $HD$ values,  generated by the Alternative models, with respect to the Null model. The top indicates the value of $n$.  On the x axis, the generative models considered, with the letter $G$ denoting  $\gamma$. }
    \label{fig:HammingDistances}
\end{figure}

In order to have an edit-based similarity/distance function as a reference point, we   use the mentioned procedure with the Hamming distance ($HD$ for short), being the most natural one to capture similarity as induced by  our generative models. Results are reported in Figure \ref{fig:HammingDistances} in terms of boxplots of $HD$ values.  We observe that, as expected,  both $PT$ and $MR$ models generate sequence pairs with an increasing trend of similarity as  $\gamma$ values increase, with a very small level of variability. This is due to the fact that the two $AM$ models produce highly similar sequences that have the same, perfectly aligned, number of common sub-sequences.   It is also to be noted that the Hamming distance confirms that  $NM$ generates  random sequences with base probabilities equal to 1/4. 
The delta values of $HD$ clearly evident from Figure \ref{fig:HammingDistances} guarantees that such a function is able to capture the increasing level of similarity between pairs of sequences produced by both $PT$ and $MR$. {\color{black}{Moreover, the perfect separation of the $HD$ value distributions between the $NM$ and the $AM$s indicates that the Hamming distance is characterised by an excellent Type I error control and power.

It is natural to ask what would happen, had we chosen an alignment-based function such as Levenshtein Distance, also known as a special case of the Needlman-Wunsch algorithm (see \cite{Gus97} for technical definitions as well as an interesting historical recollection). As shown in Figure 3 of the Supplementary Material, we can draw exactly the same conclusions for the Levenshtein Distance as for HD. The advantage of the latter over the former is that the algorithm  computing it  takes linear time and space. 
 }}

\begin{figure*}[htb] % not h only
\centering
    \includegraphics[width=0.9\textwidth]{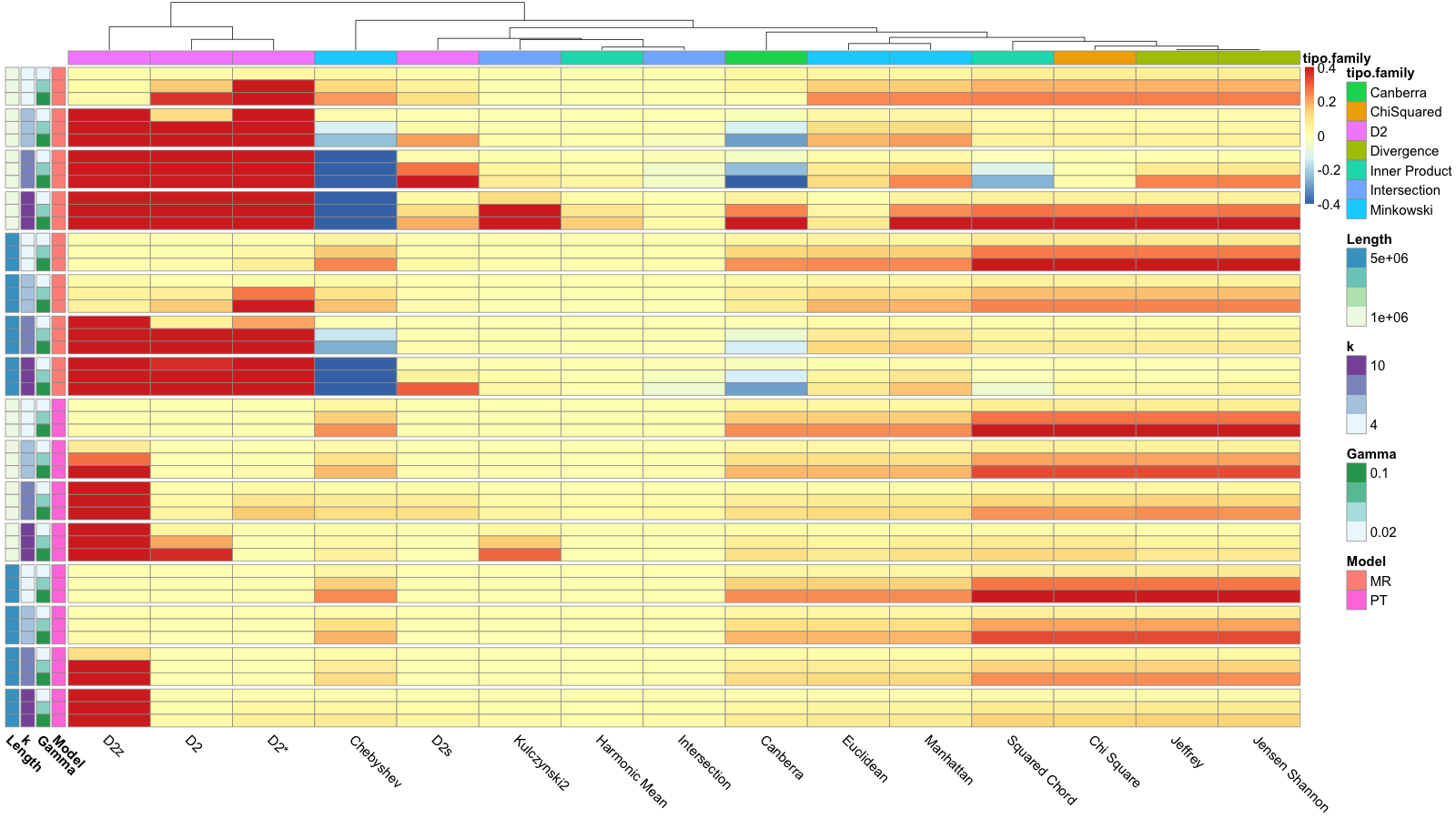}%
    \caption{Ability to capture similarity trends. Heatmap of the delta values obtained as the difference between the average of the distribution of AF values computed with   $NM$ and one of $AM$ and $PT$, for different combinations of $k$ and $\gamma$ and $n$ (see main text),  reported as colors on the left panel annotation. The dendrogram on the top is a hierarchical clustering  on the delta values with Euclidean distance and Complete Linkage. }\label{fig:Panelclustering}
\end{figure*}

Moving from edit-based {\color{black}{distances}} to {\color{black}{word-frequency}} based AF functions, we perform the same experiments with the same $\gamma$ and $n$ values,  but with the addition of  $k=4, 6, 8, 10$. The results are reported in Figure \ref{fig:Panelclustering}  as heatmap of  delta values (see Section \ref{sec:KS}). The dendrogram on the top of the figure is the result of a hierarchical clustering of AF function delta vectors using Euclidean distance and Complete Linkage clustering method \cite{jain0}. 

In order to explain the clustering,  we mention that colors in the heatmap encode the following situation, on a scale from red to blue: i) positive delta values (red colors) indicate that the AF function is able to capture the similarity with respect to the case of random sequences, i.e. higher the delta, \textcolor{black}{the better} the measure; ii) negative delta values (blue colors) indicate a complementary situation, i.e., the measure is not able to capture the similarity.

Therefore, the clustering groups functions  according to their ability to capture similarity. 

Based on all the above, Figure \ref{fig:Panelclustering} shows a global positive behaviour with the exception of Chebyshev and Canberra for the $MR$ model. Apart from those two,  the dendrogram groups AF functions into three clusters that we describe from right to left. 
\begin{itemize}
\item The right side group (Euclidean, Manhattan, SquaredChord, \ChiSquare, Jeffrey, Jensen-Shannon) behaves homogeneously well, independently of the Alternative models, but  with some dependence of $k$ and $\gamma$. That is,  for $k=4, 6$ this group captures similarity better than $k=8, 10$, as $\gamma$ increases. It is to be noted that Manhattan and Euclidean are the weakest in this group: smaller delta values (lighter red)
\item The central group (\Dds, Harmonic Mean, Intersection, Kulczynski2) is characterised by lower levels of delta, independently of the Alternative models, $k$ and $\gamma$ values.   
\item The leftmost group (\Ddstar, \Dd, \Ddz) shows the highest potential to capture similarities (highest delta values), especially for $k=8, 10$. While \Ddstar and \Dd seem to \vir{prefer} $MR$, \Ddz performs well in both Alternative models.

\end{itemize}

The previous analysis highlights AF average trends with an increasing similarity,  but no quantitative information is provided regarding the ability of an AF function to identify truly similar sequences, i.e., true positives. Therefore, we resort to  more rigorous power  tests in the next section.

\subsection{Power Estimation}\label{sec:power}

\subsubsection{Overall Classification}

\begin{figure*}[htp] % not h only
\centering
\subfigure[Pattern Transfer]{%
    \includegraphics[width=0.98\textwidth]{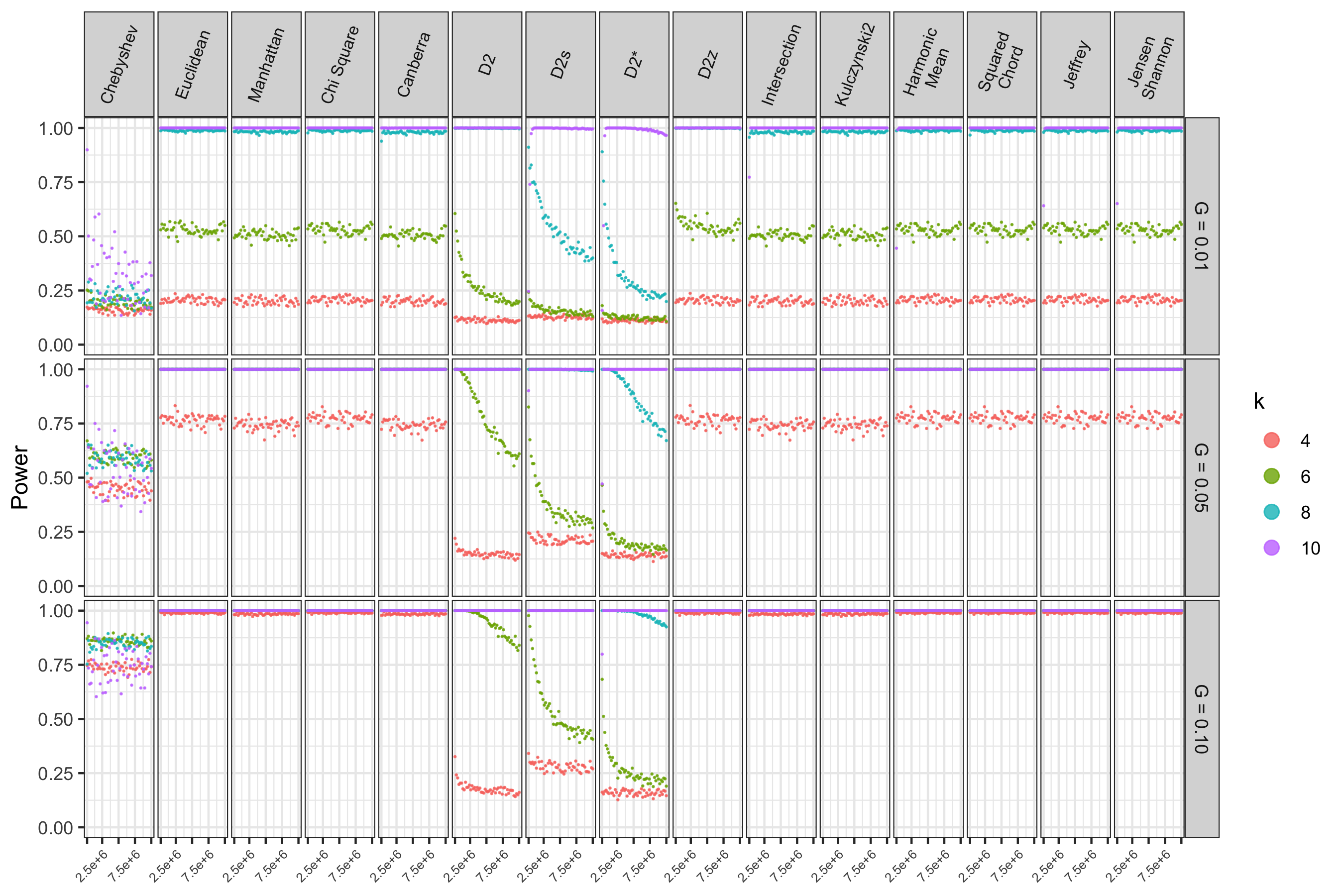}%
    }%

\subfigure[Motif Replace]{%
    \includegraphics[width=0.98\textwidth]{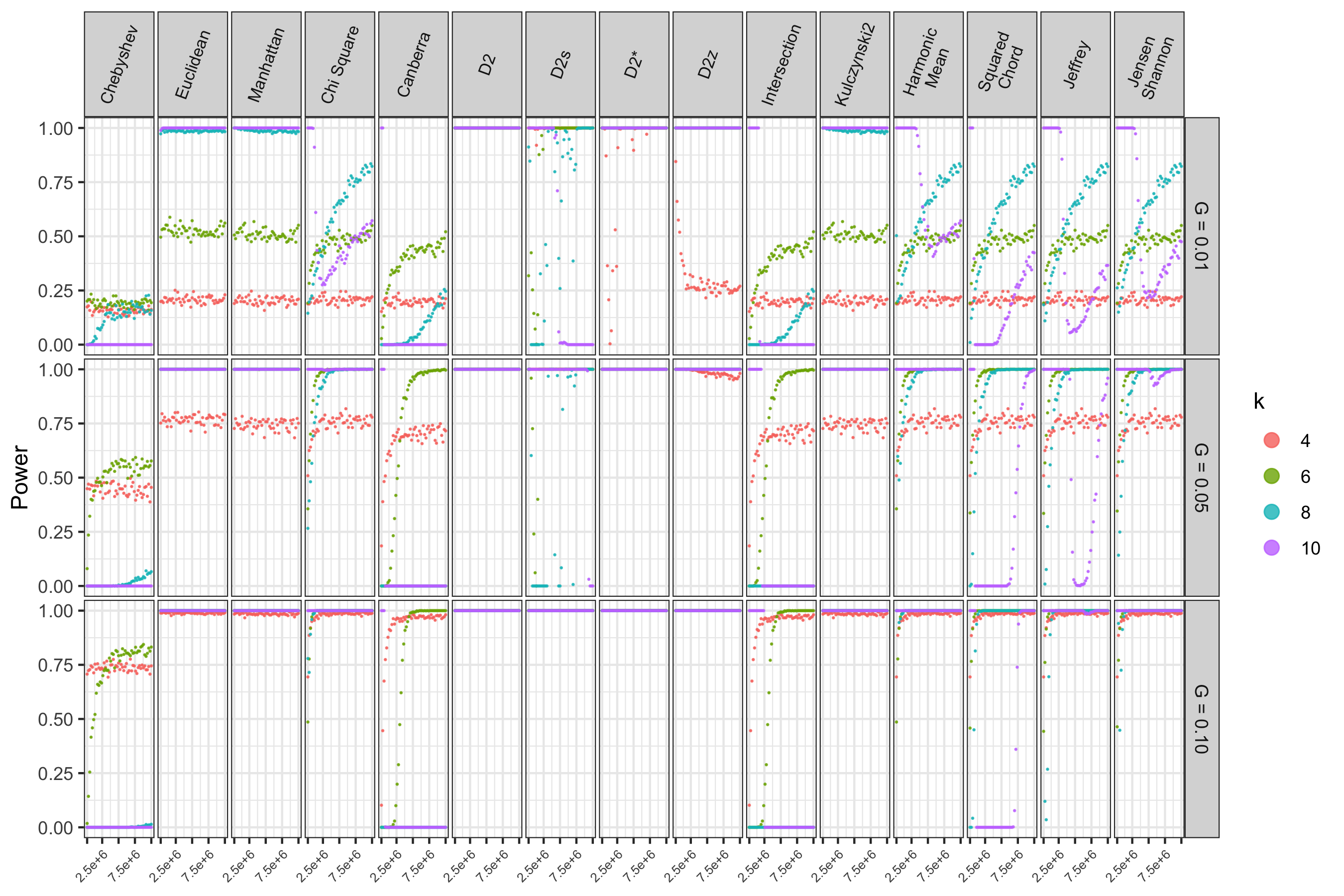}%
    }%
    \caption{Power Analyses \textcolor{black}{for the long sequences scenario}. Upper Panel: Power trend for the $PT$ Alternative model. Lower Panel: Power trend for the $MR$ Alternative model. For each AF and $\gamma=0.01,0.05,0.1$ the power levels obtained across different values of $n$ is reported and colored according to the value of $k$.}
    \label{fig:power}%
\end{figure*}

In order to evaluate statistical power of an AF measure, sequences are generated with $PT$ or $MR$  models and tested on the null distribution (obtained with the $NM$ model). The estimation has been performed for the same $n$,  $k$ and $\alpha$ as reported previously (see Section \ref{sec:resT1}),  for both the $PT$ and the $MR$ Alternative models with the addition of $\gamma=0.01, 0.05, 0.1$. 

Supplementary Figures 6 \textcolor{black}{and 7  report the distribution of the proportion of true positives,  cumulatively by length,  for respectively the long and short sequences scenario}. In agreement with the trends depicted in Figure \ref{fig:Panelclustering} for only one value of  $n$,  we find that AF power is $k$  and Alternative models dependent. In addition, it has dependence on $n$.  That is,  i) as expected with the increasing similarity between sequences (from $\gamma=0.01$ to $\gamma=0.1$),  we observe a general increase of the power; ii) $k=4$ for \textcolor{black}{short and long} $n$ ranges is generally characterised by a poor performance; iii) with some exceptions,  $k=8, 10$ show high power values \textcolor{black}{ for long sequences scenario, while worsen their performance in the case of short sequence scenario and $MR$ Alternative model};  iv) in the $MR$ model,  Canberra, Intersection, Jeffrey, Jensen-Shannon, Squared Chord and \Dds show a highly variable behaviour  in terms of $n$ and for $k=8, 10$; \textcolor{black}{v) differently form the long sequence scenario, the short one shows  highly variable power trends not only for the $MR$ but also for the $PT$ Alternative model.} Finally, the Chebyshev function is confirmed to be characterised by the worst performance in both models \textcolor{black}{and scenarios}.

\textcolor{black}{Figures \ref{fig:power} and  8 (this latter reported in the Supplementary Material)  give an even more accurate picture of the relationship between power, $k$, $n$ and $\gamma$}. We evidence some peculiar and unexpected power trends that allow us to classify AF function into two different groups: stable (when there is no functional relationship between power and $n$) and unstable (when, on the contrary, a clear increasing or decreasing trend can be observed between power and $n$). In particular,  it allows us to partition AF functions into four different classes, that we now define.

Class 1 has a stable power behaviour in both $MR$ and $PT$; class 2 has a stable power behaviour in the $PT$ but an unstable power behaviour in the $MR$;  class  3 has a stable  power behaviour in the $MR$ but an unstable power behaviour in the $PT$ and class 4 has an unstable power behaviour in both $MR$ and $PT$. Those findings are reported in \textcolor{black}{Table \ref{tab:AFclassificationLong} for both the considered scenarios}.

Unexpectedly, as reported in  Figure \ref{fig:power}, while in $MR$ model most of the functional relationships between power and $n$ show an increasing trend, in the $PT$  model most of them are characterised by a decreasing trend (for $k=6, 8$).
\textcolor{black}{Moreover,  we observe that in the short sequences scenario,  the power for $k=8, 10$ is  poor. This suggests   that these $k$ values are unsuitable for short $n$ (see again Figure 8 of the Supplementary Material). }

\ignore{
\begin{table}[]
\caption{Classification of the AF functions considered in this study according to their behavior in the $PT$ and in the $MR$ Alternative models \textcolor{black}{for short (5th column) and long (rightmost column) sequence scenario. For each model and each function, the letter U reports an unstable behavior and the letter S reports a stable behavior.}  }
\label{tab:AFclassificationLong}
\centering\begin{tabular}{clcccc}
\multicolumn{1}{c}{\textbf{AF Family}} & \multicolumn{1}{c}{\textbf{AF Method}} & \textbf{PT} & \textbf{MR} & \textbf{$Cl_{short}$} & \textbf{$Cl_{long}$}\\
\hline\\
Minkowski & Chebyshev & unstable & unstable & 4 & 4\\
Minkowski & Euclidean & stable & stable & 1 & 1 \\
Minkowski & Manhattan & stable & stable & 1 & 1\\
\ChiSquare & \ChiSquare & stable & unstable & 1 & 2 \\
Canberra & Canberra & stable & unstable & 1 & 2 \\
\Dd & \Dd & unstable & stable & 3 & 3\\
\Dd & \Dds & unstable & unstable & 4 & 4\\
\Dd & \Ddstar & unstable & stable & 3 & 3\\
\Dd & \Ddz & stable & stable & 1 & 1\\
Intersection & Intersection & stable & unstable & 3 & 2\\
Intersection & Kulczynski2 & stable & stable & 1 & 1 \\
Inner Product & Harmonic Mean & stable & unstable & 4 & 2 \\
Inner Product & Squared Chord & stable & unstable & 2 & 2 \\
Divergence & Jeffrey & stable & unstable & 4 & 2 \\
Divergence & Jensen-Shannon & stable & unstable & 4 & 2\\
\end{tabular}
\end{table}
}
\begin{table}[]
\caption{Classification of the AF functions considered in this study according to their behavior in the $PT$ and in the $MR$ Alternative models \textcolor{black}{for short and long sequence scenario. For each model and each function, the letter U reports an unstable behavior and the letter S reports a stable behavior.}  }
\label{tab:AFclassificationLong}

\centering\begin{tabular}{cl|ccc|ccc|}
\multicolumn{2}{l|}{}  & \multicolumn{3}{c|}{\textbf{Short}}  &
\multicolumn{3}{c|}{\textbf{Long}}\\
\multicolumn{1}{l}{\textbf{Family}} & \multicolumn{1}{c|}{\textbf{Method}} & \textbf{PT} & \textbf{MR} & \textbf{Class} & \textbf{PT} & \textbf{MR}& \multicolumn{1}{c|}{\textbf{Class}}\\
\hline
Minkowski & Chebyshev & U & U & 4 & U & U & 4\\
Minkowski & Euclidean & S & S & 1 & S & S & 1\\
Minkowski & Manhattan & S & S & 1 & S & S & 1\\
\ChiSquare & \ChiSquare & S & S & 1 & S & U & 2\\
Canberra & Canberra & S & S & 1 & S & U & 2\\
\Dd & \Dd & U & S & 3 & U & S & 3\\
\Dd & \Dds & U & U & 4 & U & U & 4\\
\Dd & \Ddstar & U & S & 3 & U & S & 3\\
\Dd & \Ddz  & S & S & 1 & S & S & 1\\
Intersection & Intersection & U & S & 3& S & U & 2\\
Intersection & Kulczynski2 & S & S & 1 & S & S & 1\\
Inner Product & Harmonic Mean & S & S & 1 & S & S & 1\\
Inner Product & Squared Chord & S & U  & 2 & S & U  & 2 \\
Divergence & Jeffrey  & U & U & 4 & S & U  & 2 \\
Divergence & Jensen-Shannon  & U & U & 4 & S & U  & 2 \\

\end{tabular}
\end{table}

\subsubsection{Insights into the overall classification.}
\label{subsec:insights}

%\begin{table}[]
%\caption{Classification of the AF functions considered in this study   according to their behavior in the $PT$  and in the $MR$ Alternative models \textcolor{black}{for short sequence scenario.}.  }
%\label{tab:AFclassificationShort}
%\centering\begin{tabular}{clccc}
%\multicolumn{1}{l}{\textbf{AF Family}} & \multicolumn{1}{c}{\textbf{AF Method}} & %\textbf{PT} & \textbf{MR} & \textbf{CL}\\
%\hline\\
%Minkowski & Chebyshev & unstable & unstable & 4\\
%Minkowski & Euclidean & stable & stable & 1 \\
%Minkowski & Manhattan & stable & stable & 1\\
%\ChiSquare & \ChiSquare & stable & stable & 1 \\
%Canberra & Canberra & stable & stable & 1 \\
%\Dd & \Dd & unstable & stable & 3\\
%\Dd & \Dds & unstable & unstable & 4\\
%\Dd & \Ddstar & unstable & stable & 3\\
%\Dd & \Ddz & stable & stable & 1\\
%Intersection & Intersection & unstable & stable  & 3\\
%Intersection & Kulczynski2 & stable & stable & 1 \\
%Inner Product & Harmonic Mean & unstable & unstable & 4 \\
%Inner Product & Squared Chord & stable & unstable & 2 \\
%Divergence & Jeffrey & unstable & unstable & 4 \\
%Divergence & Jensen-Shannon & unstable & unstable & 4\\
%\end{tabular}
%\end{table}

%The power statistics experiments have lead to a useful 
%classification of AF functions in regard to their ability to  capture the similarity of two sequences, this latter as formalized by two different models. 
\textcolor{black}{Looking at the AF classes reported in Table \ref{tab:AFclassificationLong}, it is clearly evident that $n$ affects AF classification at least for some methods. In particular \ChiSquare, Canberra, Intersection, Harmonic Mean, Jeffrey and Jensen-Shannon methods show different power trends for different $n$ ranges: while \ChiSquare and Canberra seem to become more stable moving from long to short sequences, Intersection, Harmonic Mean, Jeffrey and Jensen-Shannon seem to become more unstable. Minkowski and \Dd families seem to behave consistently across long and short sequence scenarios.}

Apart from AF functions belonging to class 4, the previous classification evidences that some  AF functions  previously characterized by a reasonable good ability to capture similarity trend (see Section \ref{sec:capture}), to a closer and more rigorous scrutiny show poor power in, at least, one model. In order to provide a reason for such a behaviour,  we provide an additional and more detailed analysis of four functions, representative of differing behaviours in terms of power. They are:  \Ddz for class 1, \ChiSquare and  Intersection for class 2 \textcolor{black}{in the long sequence scenario)}, and \Ddstar for class 3. We take as reference the power results in Figure \ref{fig:power}.

\begin{figure*}[htp] % not h only
\centering
\subfigure[\ChiSquare]{%
    \includegraphics[width=0.23\textwidth]{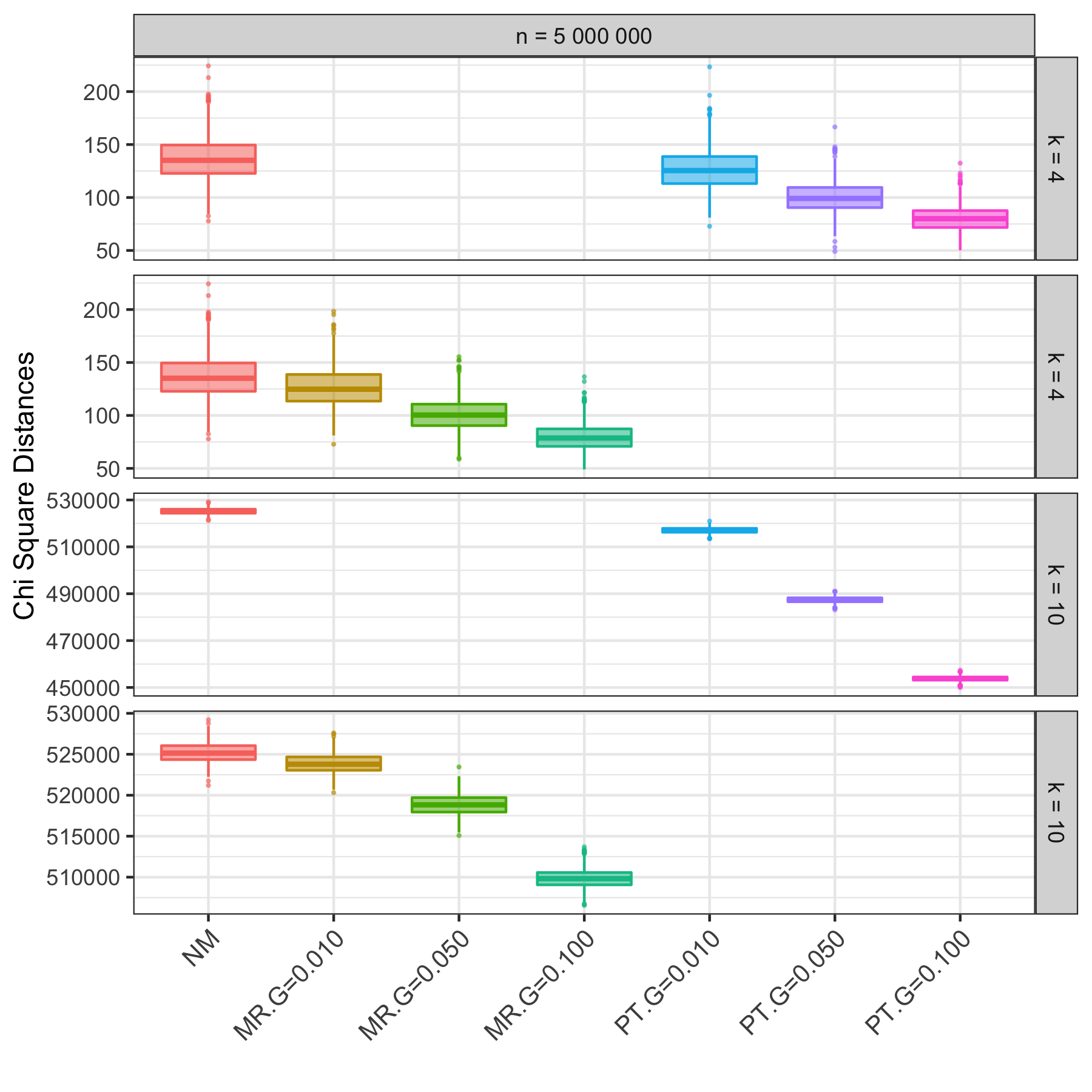}%
    }%
\hfill
\subfigure[intersection]{%
    \includegraphics[width=0.23\textwidth]{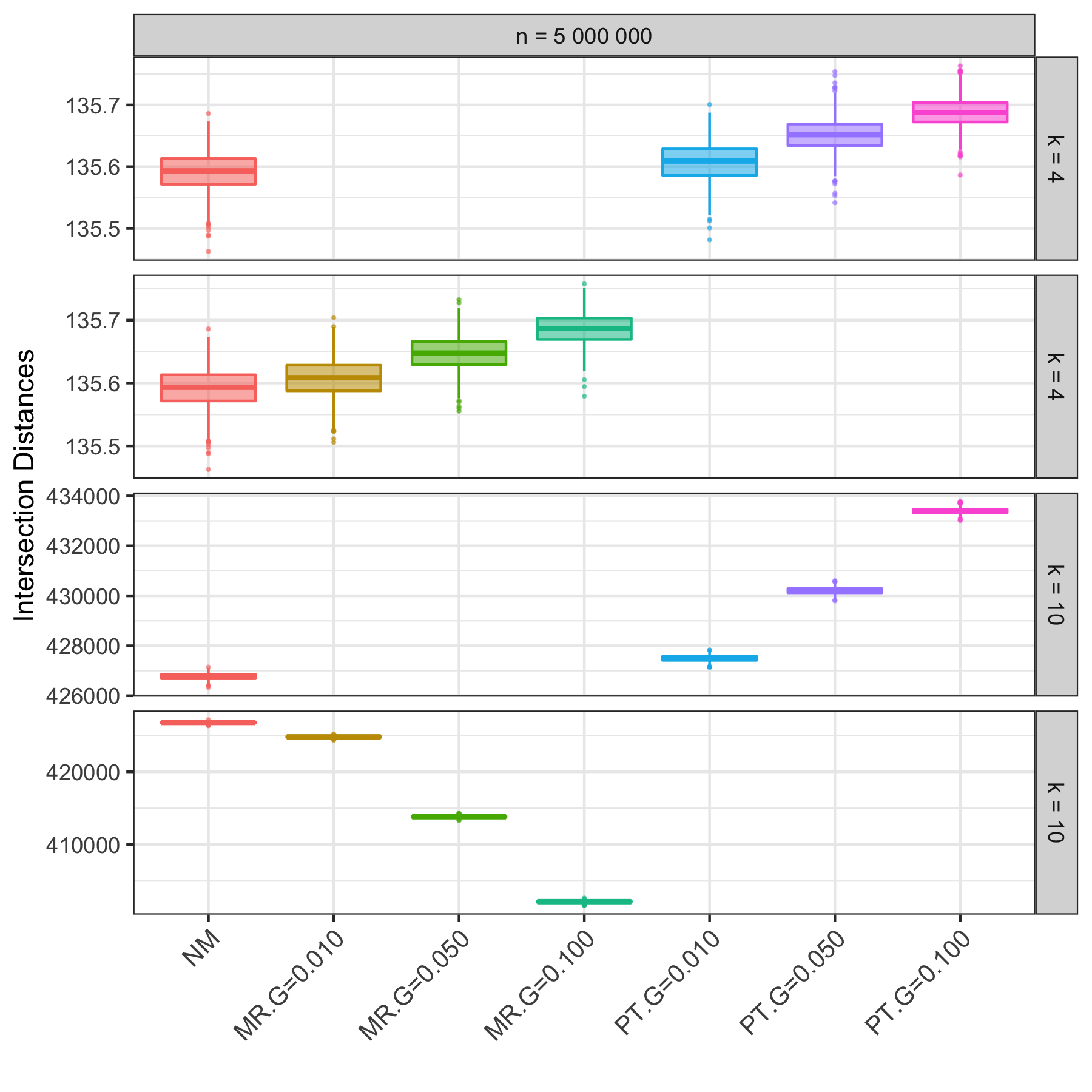}%
    }%
\hfill
\subfigure[\Ddstar]{%
    \includegraphics[width=0.23\textwidth]{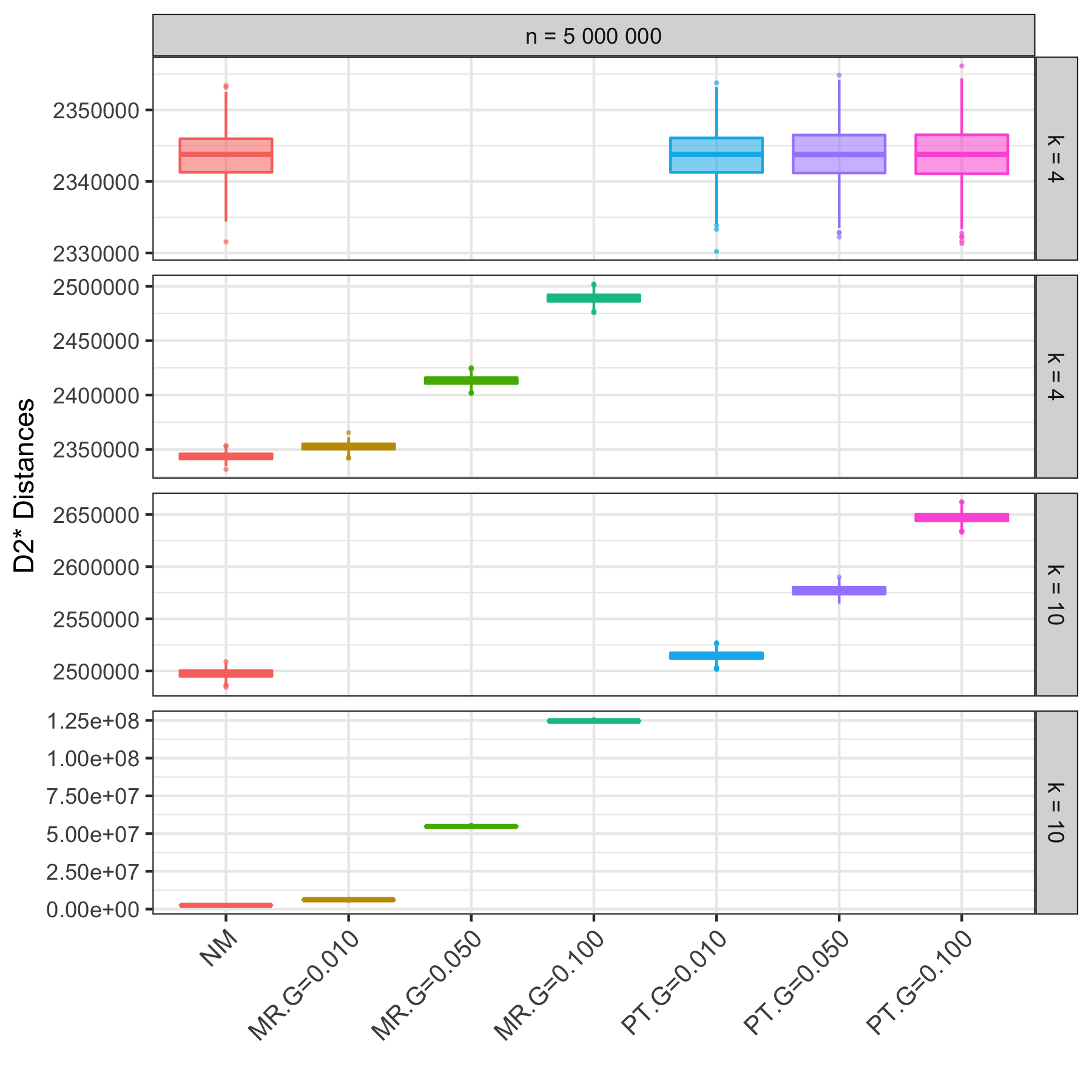}%
    }%
\hfill
\subfigure[\Ddz]{%
    \includegraphics[width=0.23\textwidth]{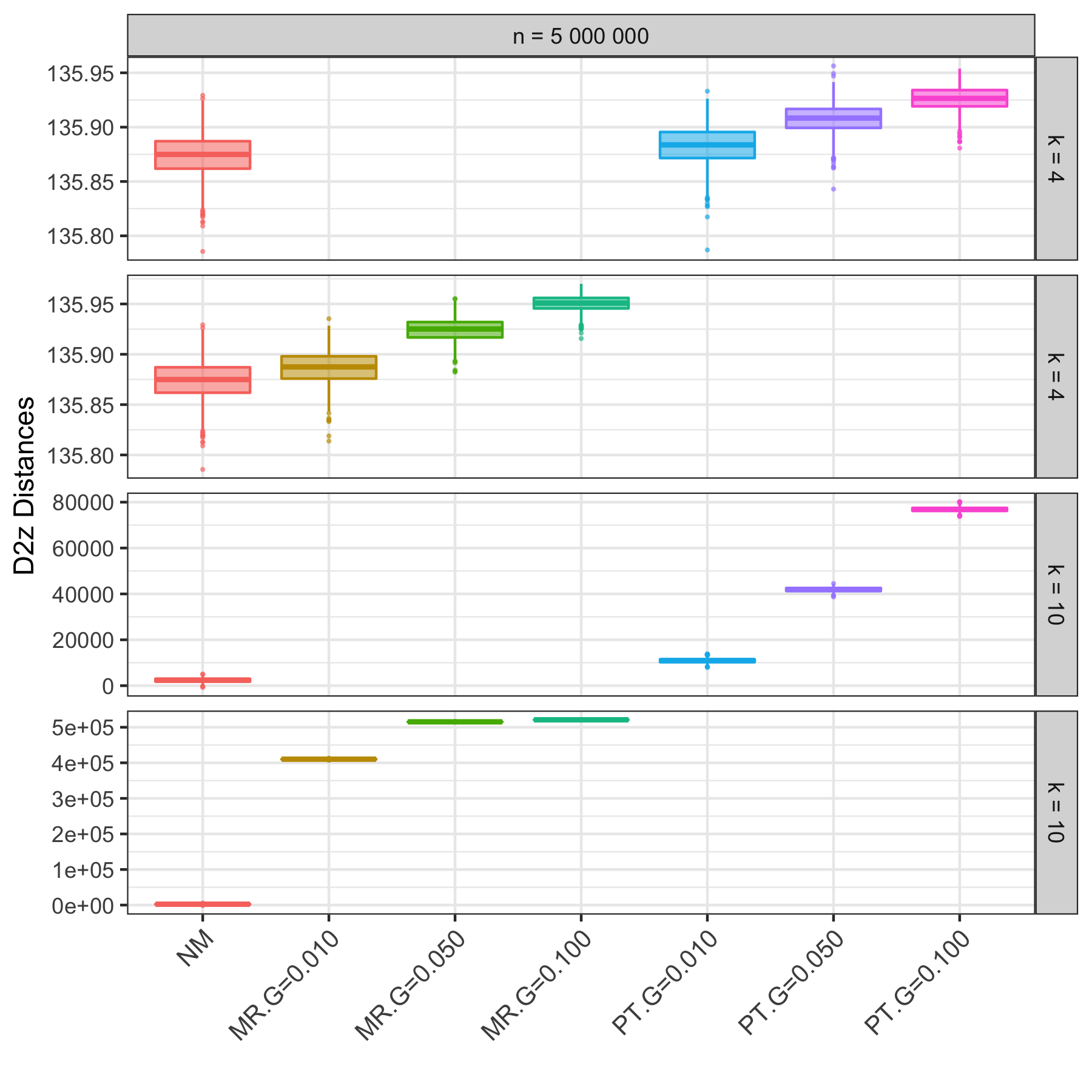}%
    }%
\caption{Separation Power results for (a) \ChiSquare (b) Intersection (c) \Ddstar (d) \Ddz. For each AF function, each panel corresponds to a different value of $k$ (MR at bottom and PT at top). On the abscissa, the generative models considered, with the letter $G$ denoting  $\gamma$.  For each experiment, the boxplot of the $HD$ values computed on the corresponding set of pairs of sequences is reported on the ordinate.}
%The remaining part of the legend is as in Figure~\ref{fig:PlotChisquareDistances}.}%
\label{fig:PlotChisquareDistances}
\end{figure*}

In the $MR$ model, Intersection  shows an opposite behaviour than the expected one for $k=10$: it has an optimal power for small values of $n$ but it collapses to zero for $n$ greater than $2\ 000\ 000$. This drop is very well explained and exemplified by the  counter-intuitive decreasing trend of the function values   for $k=10$, as it is clearly visible in the boxplots of Figure \ref{fig:PlotChisquareDistances}(b). Keeping in mind that Intersection is a similarity measure while $HD$ is a distance, for the $MR$, it is evident the difference in value  distributions of the former with respect to the latter. Indeed, in Figure \ref{fig:PlotChisquareDistances}(b) (MR), the boxplots of the former have tendency opposite to the one that a similarity measure should have, while in Figure \ref{fig:HammingDistances}, the boxplots have the correct tendency and are very well separated.

In the $MR$ model, \ChiSquare shows a poor power for $k=4$ independently from $n$ while, for higher $k$s, a better trend is evident. This is confirmed  in Figure~\ref{fig:PlotChisquareDistances}(a) by the overlapping boxplots shown  for $k=4$.  As in the case of 
Intersection, it is useful to compare those  boxplot trends with the ones of $HD$ in  Figure~\ref{fig:HammingDistances}.

\Ddstar exhibits an opposite behaviour with respect to the other functions: it reaches a perfect power in the $MR$ model independently from $k$ and $n$,  while it has poor power in the $PT$ model for $k\leq 10$. This is explainable by the completely overlapping boxplots  for $k=4$ in Figure~\ref{fig:PlotChisquareDistances}(c). Again, it is useful to compare those  boxplot trends with the ones of $HD$ in  Figure~\ref{fig:HammingDistances}. 

\Ddz is the only function that exhibits a 100\% of power for $k=8, 10$, independently from $n$ and from the Alternative models, as confirmed in Figure~\ref{fig:PlotChisquareDistances}(d). Indeed, those boxplots behave in complete analogy with the ones of 
$HD$ displayed in  Figure~\ref{fig:HammingDistances}.

In conclusion, poor power or a sudden drop in power can be explained as a loss of the ability of an AF function to produce values in sequence pairs generated with the Alternative models clearly disjointed from the ones produced in the case of Null model, with an increasing or decreasing trend in the case of a similarity or distance function, respectively.

\section{Discussion}
Our experiments, together with the analysis of the results, provide a comprehensive study of the  {\color{black}{word-frequency} based} AF function performance in terms of Type I error control and power using synthetic datasets.  Although empirical evaluations on gold standard real datasets are certainly  appropriate for Type I error control evaluation, they are challenging in the case of power analyses as true positives (real sequence similarity) are hard to be defined. 

%that suggests validation procedures of novel AF function different than the ones being now followed in the Literature. Indeed, empirical evaluation on gold standard datasets is certainly still appropriate. However, limiting the validation of a new AF function to Type I error control only is not informative, when using synthetic datasets simulating biological processes. 

Our results demonstrate that, although most of  the tested AF functions perform well in controlling false positives, much more compelling are the results obtained in the power studies, showing different and unexpected performances. 

One of the reasons for the scarce use of comparative power studies is the absence of effective computational tools. Therefore, as an additional contribution to the advancement of the State of the Art, we also provide an entire software system, based on Big Data Technologies, to rapidly carry out  Power experiments. 

Moreover, we get the following additional more detailed insights.

\begin{itemize}
    \item [(1)] {\bf Type I Error Control and the Choice of $k$.} The good performance of the selected AF functions in this setting is somewhat expected. %Indeed, pragmatically, if those functions would abound in returning \vir{false positives} then the number of studies concentrating on them would be largely unjustified. 
    However, using a rigorous,  coherent and quantitative evaluation  of the proportion of false positives, we show that this proportion is not much affected by the chosen value of $k$,  a  novel and somewhat unexpected result. 
    
    %Indeed, a well known and much used heuristic for the choice of $k$  (see \cite{luczak2017survey,Zielezinski2019}) is to pick a value   logarithmic in the sequence lengths. Our sequences have very differing lengths and, for each, we have used the mentioned values of $k$, independently of considerations  regarding length. 
    
    \item[(2)] {\bf Beyond a Syntactic Grouping of AF Functions by \vir{Family}.} 
    Our experiments show the presence of four main AF function groups that substantially differ from the syntactic grouping provided by \cite{luczak2017survey} (Figure 1, but see also \cite{Bernard16}). We further observe that  there is enough   \vir{information} regarding similarity  in the $k$-mer statistics used by AF functions. Indeed, an alignment method such as  $HD$ and  a {\color{black}{word-frequency} based} one such as   \Ddz have essentially the same performance. Rather, the poor performance of some measures seems to critically depend, in a non-obvious way, from the formula characterizing the function. Indeed,  AF functions that are mathematically closely related, e.g., Manhattan and Canberra, may exhibit quite different power. 
    
    \item[(3)] {\bf The Complex and  Function Specific Dependency of Power on Key AF Functions Parameters.} Based on our experiments, the ability to detect true positive similarities is deeply dependent on the value of $k$, the length of the sequences $n$ and the biological reality that the generative models are meant to encode. In particular, we note a poor power for some AF functions,  even when $k$ is selected using the already mentioned time-honored heuristic to choose $k$ as the log of sequence length (i.e. $k=4, 6$ for short sequences). In details, for Type I Error, the relationship between $k$ and $n$ seems to be irrelevant (see (1) above), while here it is AF function specific. Those findings indicate that the identification of a rule for the selection of $k$, even heuristic, general to all AF functions can be a  very evanescent task. In this respect, our data analysis  is useful to empirically guide the selection of $k$ according  to the sequence length $n$.

    \item[(4)] {\bf The \Dd Family: Its Power in Relation to Alternative Models.}
    This is a very prominent family  in Alignment-Free  Computational Biology \cite{song2013new}. Relevant for this research is the fact that it is known  that \Ddstar and  \Dd do not perform adequately on the $PT$ model \cite{2009reinertalignment,wan2010alignment}. Variants as a remedy to this  have  been proposed
\cite{Liu2011},  but due to their computational  cost,  they are mainly of theoretic interest. We add some new facts regarding this important family. First, all of the \Dd family performs well on both models, across sequence length, but for \vir{large} $k$s. 
Moreover, \Ddz is the best performer on both $PT$ and $MR$, for all values of $k$ we have tested. 
\end{itemize}

\textcolor{black}{
\section{Conclusions and Open Problems}}
In this study, we have provided a statistically sound and comprehensive study of {\color{black}{word-frequency} based} AF functions in terms of Type I error  control and power. Our findings indicate the need  to validate new AF functions via Power rather than Type I error control ones, as done so far. In this respect, we also offer software based on Big Data Technologies that makes rather easy the study of new {\color{black}{word-frequency}} based AF functions in terms of Power. We also provide guidelines on the choice of $k$, a key parameter, that turns out to be function-dependent rather than \vir{universal}. Finally, we also identify an AF function, i.e. the \Ddz, that performs very well, across values of $k$, sequence lengths, and Alternative models. Despite the \Dd family, one of the most prominent AF families,  has been the object of many investigations, the excellence of \Ddz in terms of power has not been reported in previous studies. 

{\color{black}Furthermore, this research points out the need for further studies in this area. The first asks for \vir{power studies}, with models in which the motifs have gaps and mismatch rather than being exact copies. The second is related to the identification of proper theoretic and experimental settings in which to study  the power of  absent/present word based AF functions. Indeed, important functions such as Jaccard and its fast approximation Mash are part of this family, but even a quite established methodology, such as the one by Reinert et. al, is inadequate to characterize their power. Finally, it is very important to extend this type of study to additional families of AF functions, in particular the ones based on micro-alignments (see \cite{Leimester17} and references therein).
}

\section*{Acknowledgements}
All authors would like to thank the GARR Consortium for having made available a cutting edge OpenStack Virtual Datacenter for this research. 

\section*{Funding}
G.C., R.G. and U.F.P. are partially supported by GNCS Project 2019 \vir{Innovative methods for the solution of medical and biological big data}. R.G. is also supported by  MIUR-PRIN project \vir{Multicriteria Data Structures and Algorithms: from compressed to learned indexes, and beyond} n. 2017WR7SHH. 
U.F.P. and F.P. are partially supported by Universit\`{a} di Roma - La Sapienza Research Project 2020 \vir{Algoritmi su grafi, limitazioni nel sequenziale e opportunità nel distribuito}.
C.R. is supported by the Italian Association of Cancer Research (AIRC) (n. IG21837). 

\bibliographystyle{abbrv}



\section*{Supplementary material}

# The Power of Word-Frequency Based Alignment-Free Functions: a Comprehensive Large-scale Experimental Analysis - Version 3
## Supplementary Material

Giuseppe Cattaneo*[†]    Umberto Ferraro Petrillo[‡]    Raffaele Giancarlo[§]

Francesco Palini[‡]    Chiara Romualdi[¶]

**Abstract**

Additional details about the Main Manuscript are provided in this document.

# 1 Distance Functions

In this section, we provide definitions of the AF functions included in this study. In what follows, we adopt both the classification and the notation from [1]. It is to be remarked that some of the AF functions defined next appear in the Literature with different names or they are easy variants of the functions introduced here. For instance, FFP adopted in [4] is the well known Jensen-Shannon Divergence defined here. For the interested reader, the original publication where the functions defined here have been introduced can be found in [1] for the less known cases.

Given a set of sequences $S = \{s_1, ..., s_n\}$, a $k$-mer statistics $h_s$ for a sequence $s$ in the set is defined as follows:

$$h_s = \langle c(w_1), c(w_2), ..., c(w_{|K|})\rangle \tag{1}$$

where $c(w_i)$ is the number of occurrences of the word $w_i$ (i.e. the $i$-th $k$-mer) in the sequence $s$ and $K$ is the set of all possible words of length $k$ over the alphabet $\{A, C, G, T\}$.

## 1.1 The Minkowski Family

Given two sequences $s$ and $t$ and their associated statistics $h_s$ and $h_t$, the *Euclidean* distance is defined as:

$$Euclidean(h_s, h_t) = \sqrt{\sum_{w \in K} (h_s(w) - h_t(w))^2} \tag{2}$$

A widely adopted variant of the *Euclidean* distance is the *Manhattan* distance:

$$Manhattan(h_s, h_t) = \sum_{w \in K} |h_s(w) - h_t(w)| \tag{3}$$

Another member of this family, the *Chebyshev* distance, is based on the idea of applying the $p$-th root on the *Manhattan* distance, with $p \to \infty$, and it is defined as follows:

$$Chebyshev(h_s, h_t) = \max_{w \in K} |h_s(w) - h_t(w)| \tag{4}$$

---

*Dipartimento di Informatica, Università di Salerno, Fisciano (SA), 84084, Italy

[†]To whom correspondence should be addressed.

[‡]Dipartimento di Scienze Statistiche, Università di Roma - La Sapienza, Rome, 00185, Italy

[§]Dipartimento di Matematica ed Informatica, Università di Palermo, Palermo, 90133, Italy

[¶]Dipartimento di Biologia, Università di Padova, Padova, 35131, Italy.

## 1.2 The $\chi^2$ Distance

It is defined as:

$$\chi^2(h_s, h_t) = \sum_{w \in K} \frac{(h_s(w) - h_t(w))^2}{(h_s(w) + h_t(w))} \quad (5)$$

## 1.3 The Canberra Distance

It is a mix between Manhattan and $\chi^2$ distances:

$$Canberra(h_s, h_t) = \sum_{w \in K} \frac{|h_s(w) - h_t(w)|}{(h_s(w) + h_t(w))} \quad (6)$$

## 1.4 The $D_2$ Statistics

It expresses the similarity of two sequences in a very natural way as a sort of inner product between two statistics as shown in equation (7).

$$D_2(h_s, h_t) = \sum_{w \in K} h_s(w) h_t(w) \quad (7)$$

However, extensive studies of this function suggest that a standardized version of it is more useful in the AF sequence analysis setting. Such a variant is denoted $D_2^Z$. It is as $D_2$, except that the $k$-mer statistics have been standardized via the well known z-score transformation. Details can be found in [1].

We also consider two other members of the $D_2$ family, denoted $D_2^S$ and $D_2^*$ and described in [5].

$D_2^S$ is based on the finding [3] that if two independent random variables $X$ and $Y$ are normally distributed with mean zero, also $\frac{XY}{\sqrt{X^2+Y^2}}$ is normally distributed. We normalize $h_s(w)$ and $h_t(w)$ as follow:

$$\tilde{h}_s(w) = h_s(w) - np_s(w)$$

and

$$\tilde{h}_t(w) = h_t(w) - mp_t(w)$$

In which $n$ and $m$ are the number of $k$-mers, respectively, in $S$ and $T$, while $p_s(w)$ and $p_t(w)$ are the probabilities of the $k$-mer $w$ under the background model for, respectively, $S$ and $T$.

The $D_2^S$ statistics is defined as follow:

$$D_2^S = \sum_{w \in K} \frac{\tilde{h}_s(w) \tilde{h}_s(w)}{\sqrt{\tilde{h}_s(w)^2 + \tilde{h}_s(w)^2}} \quad (8)$$

The $D_2^*$ statistic is based on the idea that, for relatively long $k$-mers, the number of occurrences can be approximated by a Poisson distribution, thus mean and variance are nearly the same.

The $D_2^*$ statistics is defined as follow:

$$D_2^* = \sum_{w \in K} \frac{\tilde{h}_s(w) \tilde{h}_s(w)}{\sqrt{np_s(w) mp_t(w)}} \quad (9)$$

## 1.5 The Intersection Family

From this family, we selected two distances. The first is the *Intersection* distance, also known as *Czekanowski*, is based on the intersection of the $k$-mers counts divided by their union. It is:

$$Intersection(h_s, h_t) = \sum_{w \in K} \frac{2 * min(h_s(w), h_t(w))}{h_s(w) + h_t(w)} \quad (10)$$

The second is the *Kulczynski2* distance;

$$Kulczynski2(h_s, h_t) = A_\mu \sum_{w \in K} min(h_s(w), h_t(w)) \qquad (11)$$

where $A_\mu$ is equal to $\frac{4^k(\mu_s - \mu_t)}{2\mu_s \mu_t}$ ($\mu_s$ and $\mu_t$ denote the mean of the statistics $h_s$ and $h_t$, respectively).

## 1.6 The Inner Product Family

The *Harmonic Mean* distance is:

$$HarmonicMean(h_s, h_t) = 2 \sum_{w \in K} \frac{h_s(w) h_t(w)}{h_s(w) + h_t(w)} \qquad (12)$$

As opposed to the Euclidean distance that computes the square root over the summation value, the *Squared Chord* computes the square root over each value of the statistics independently:.

$$SquaredChord(h_s, h_t) = \sum_{w \in K} \left(\sqrt{h_s(w)} - \sqrt{h_t(w)}\right)^2 \qquad (13)$$

This can be simplified as

$$= \sum_{w \in K} h_s(w) + h_t(w) - 2\sqrt{h_s(w) h_t(w)} \qquad (14)$$

## 1.7 The Divergence Family

This family uses probabilities to compare two sequences measuring the distance in the log-probability space. From such a family, we selected the *Jeffrey* and the *Jensen-Shannon* distances. The former is defined as:

$$Jeffrey(h_s, h_t) = \sum_{w \in K} (p_s(w) - p_t(w)) \ln \frac{p_s(w)}{p_t(w)} \qquad (15)$$

while the second (JSD for short) is defined in 16:

$$\begin{aligned} JSD(h_s, h_t) = &\frac{1}{2} \sum_{w \in K} p_s(w) \log_2 \left(\frac{p_s(w)}{\frac{1}{2}(p_s(w) + p_t(w))}\right) \\ &+ \frac{1}{2} \sum_{w \in K} p_t(w) \log_2 \left(\frac{p_t(w)}{\frac{1}{2}(p_s(w) + p_t(w))}\right) \end{aligned} \qquad (16)$$

where $p_s(w)$ is the empirical probability of $k$-mer $w$ over all the strings of length $k$ from the alphabet $\{A, C, G, T\}$ in the input sequence $s$ and therefore $p_s(w) = \frac{h_s(w)}{4^k}$.

# 2 Generative models for sequences: technical details

## 2.1 Pattern Transfer

Fix an integer $\ell$, whose role will be clear immediately. First, $m$ pairs of sequences are generated via $Multinom(p, n)$. Then, each pair $(x, y)$ is made more similar as follows. For each position $i$ of $x$ (with $i = 1 \ldots n$), we randomly sample from a Bernoulli distribution $Bern(\gamma)$ ($\gamma$ success probability). In case of success at position $i$, the sub-sequence $x[i, i + \ell - 1]$ replaces the sub-sequence $y[i, i + \ell - 1]$ and the next sampled position is $i + \ell$ (the sequence that has been replaced is hereafter called motif and $\ell$ denotes its length).

The result is a pair of sequences $(x, \hat{y})$, where $\hat{y}$ is intuitively more similar than $y$ to $x$. Such a similarity increases with the increase of the probability of success $\gamma$. Finally, the set of the new $m$ pairs of $(x, \hat{y})$ is returned as output. We refer to this model as $PT(\ell, \gamma)$. Again, when it is clear from the context, we omit mention of its parameters.

## 2.2 Motif Replace

First, $m$ pairs of sequences of length $n$ are generated as in the case of $PT$. Second, we build a set $M$ of $d$ distinct motifs, each of length $\ell$ using a simple variant of $NM$ (details left to the reader). The number $d$ is chosen to respect the proportion used in [2] regarding the number of motifs available for insertion into a pair of sequences, with respect to their length. In that case, a single motif of length 5 for sequences of length up to $\hat{m} = 25\,000$ was used. Accordingly, taking as reference the maximum sequence length used in Reinert et al., we choose $d = \frac{n}{\hat{m}\ell}$. Third, when a position $i$ in a pair of random sequences $(x, y)$ is chosen for replacement, a motif $t$ is randomly sampled from $M$ (with replacement) and it is copied in both $x[i, i+\ell-1]$ and $y[i, i+\ell-1]$. We refer to this model as $MR(d, \ell, \gamma)$. Again, when it is clear from the context, we omit mention of its parameters.

## 3 Alternative vs Null Models: the Ability to Capture Similarity Trends, technical details

Given $NM$ and $PT$ with parameter $\gamma$, the method we propose for the quantification of how well a function , in terms of its value, separates "truly" similar sequences from those that are not follows:

(1) Generate a set $S$ of $m$ pairs of sequences, each of length $n$, via $NM$. Using $S$, generate a set $S_\gamma$ of $m$ pairs of sequences, each of length $n$, via $PT$ with parameter $\gamma$.

(2) Compute $D$ for each pair of sequences in $S$ and $S_\gamma$, respectively, and in agreement with the function parameters, e.g., $k$.

(3) $AF$ values computed in (2) are reported as boxplots and also as delta values, i.e differences between $AM$ and $NM$ average values, those latter in the form of heatmaps.

It is to be pointed out that $S_\gamma$ can be computed with different values of $\gamma$. It is expected that a good similarity function would return large delta and non-overlapping boxplots, with the ones of $S_\gamma$ "higher" than those of $S$.

## 4 Type I Error Control and Statistical Power

Given our generative models, $NM$ is used to define the null distribution, while $PT$ and $MR$ are used to generate pairs of similar sequences (the true positive pairs). For each pair of sequences generated by $NM$, $PT$ or $MR$, we estimate the number of sequence pairs for which a given AF function has a value more extreme than the percentile $1 - \alpha$ (with $\alpha$ set to $0.01, 0.05, 0.1$) of the null distribution (that is, we reject the $H_0$). If sequence pairs are generated with the $NM$ model (sequences are similar by chance) the rejection of $H_0$ identifies a false positive, otherwise if sequence pairs are generated with $AM$, we identify a true positive.

Under $H_0$, with a nominal level of $\alpha$, the expected fraction of false positives obtained by chance is $\alpha$. Then, a statistical test with a reasonably good control of Type I error, has a false positives rate close to the nominal level $\alpha$.

On the other hand, the power of a statistical test is defined as the probability of correctly rejecting $H_0$, i.e. a true positive is found. That is, letting $\beta$ be the probability of not rejecting $H_0$ when it is false, the power of a test statistic is $1 - \beta$.

Here, to evaluate the statistical power of an AF measure, sequences are generated separately with $PT$ or $MR$ and tested on the null distribution, via $NM$. In this case, higher the number of sequence pairs identified as significantly similar is, higher the power.

## 5 Results: Additional Material

### 5.1 Type I error control: additional figures

Here we report additional figures with respect to Section 3.1 of the main text, i.e., Figure 1 and Figure 2.

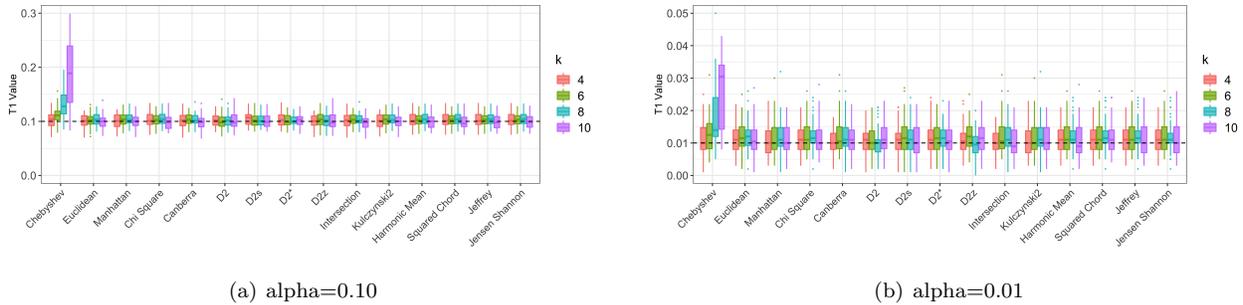

(a) alpha=0.10

(b) alpha=0.01

Figure 1: Results of the T1 Error Control experiments, for the long sequences scenario, for nominal levels 0.10 (left panel) and 0.01 (right panel). The dotted black line corresponds to the nominal level.

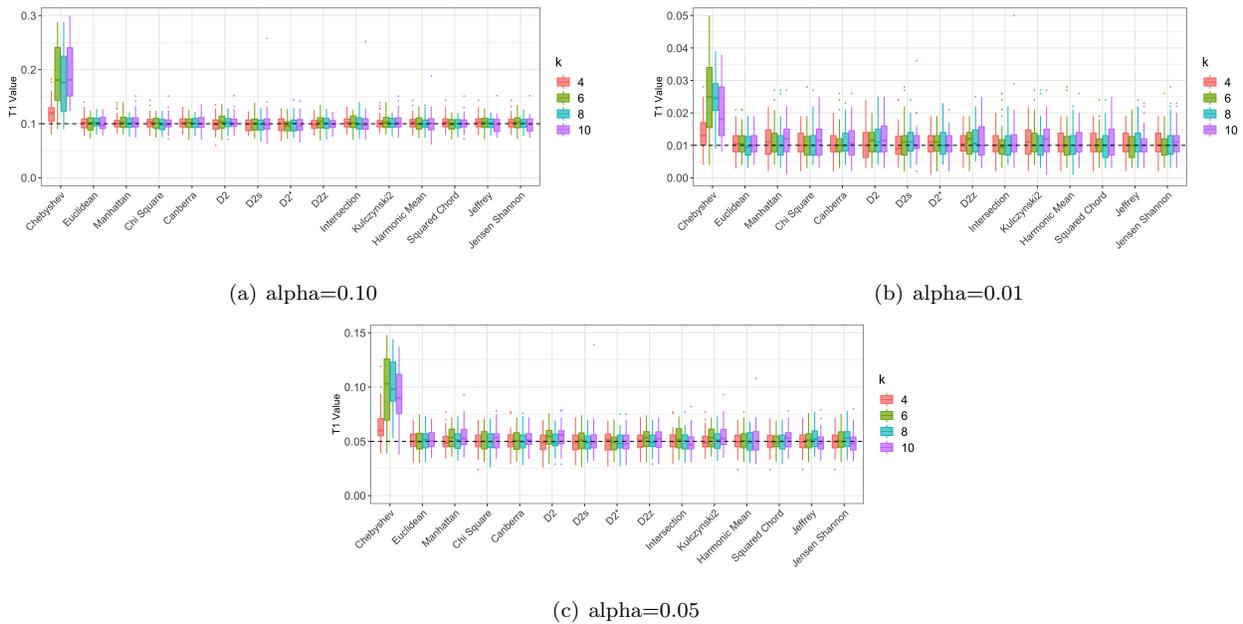

(a) alpha=0.10

(b) alpha=0.01

(c) alpha=0.05

Figure 2: Results of the T1 Error Control experiments, for the short sequences scenario, for nominal levels 0.10 (a), 0.01 (b) and 0.05 (c). The dotted black line corresponds to the nominal level.

## 5.2 Ability to Capture Similarity Trends: The Levenshtein Distance

We report in Figure 3 the boxplots of Levenshtein Distance values, providing a graphical view of the distance of values generated by the Alternative models, with respect to the Null model. The behavior we observe is the same as the one reported for HD values and described in Section 3.2 of the Main text.

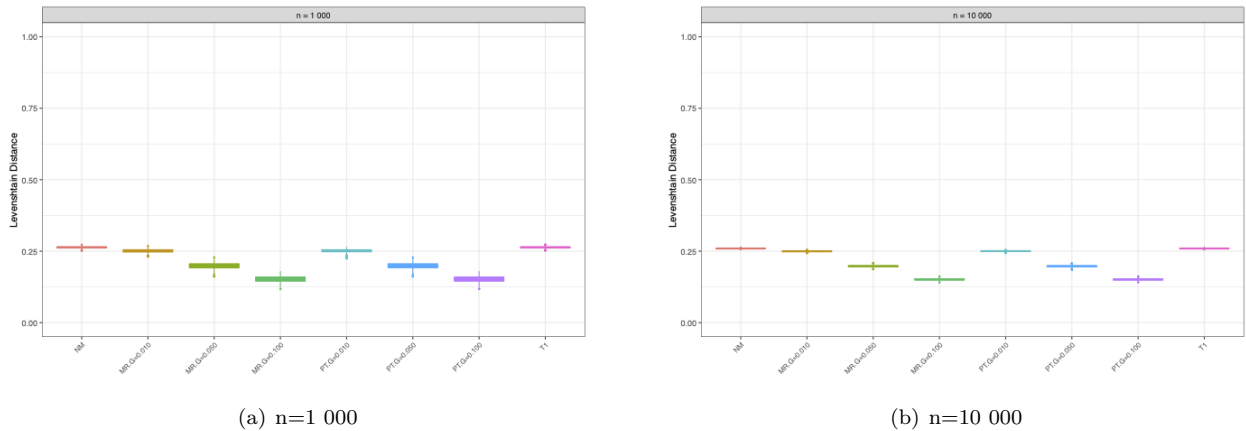

(a) n=1 000

(b) n=10 000

Figure 3: Separation, graphical view, of the Levenshtein Distance values for $n = 1\ 000$ (left panel) and $n = 10\ 000$ (right panel). The remaining part of the legend is as in Figure 2 of the Main Text.

## 5.3 Ability to Capture Similarity Trends: additional figures

Here we report additional figures with respect to Section 3.2, i.e., Figures 4 - 5.

## 5.4 Power Estimation: additional figures

Here we report an additional figure with respect to Section 3.3, i.e., Figure 6 and Figure 7.

## 5.5 Insights into the power of AF functions: additional figures

Here we report additional figures with respect to Section 3.3.2, i.e., Figures 4 - 5.

# Acknowledgements

# Funding

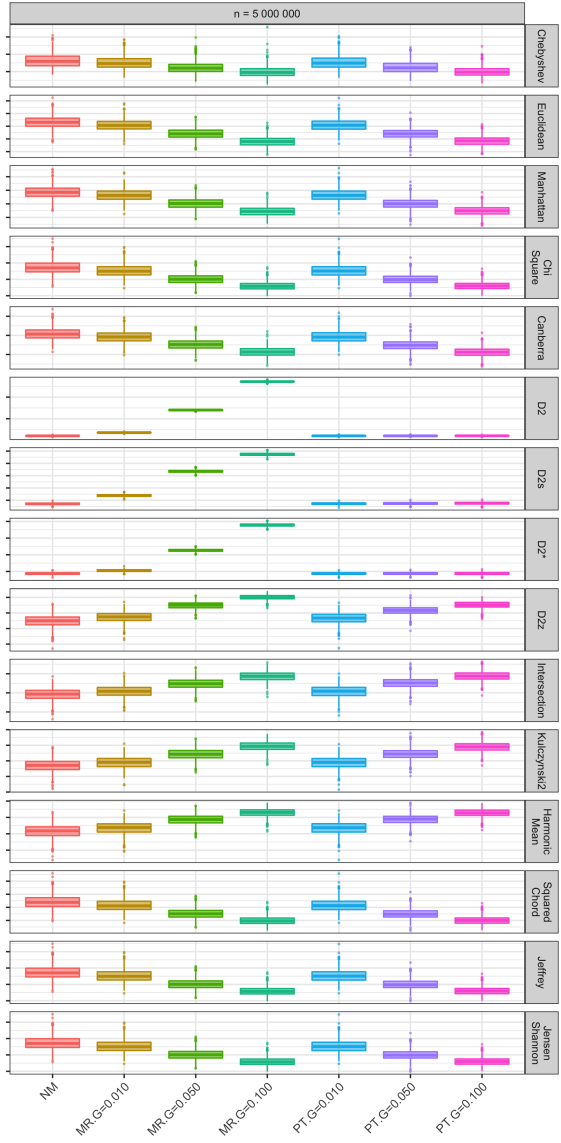
(a) k=4

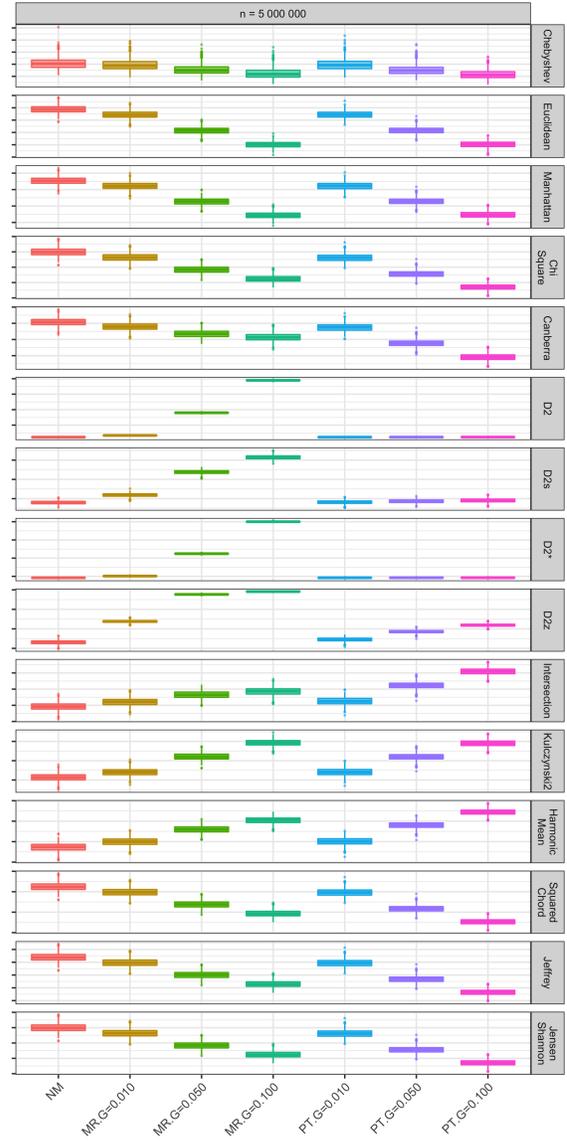
(b) k=6

Figure 4: Separation, graphical view, of the AF function values for all measures and for $n = 5\,000\,000$ and $k = 4$ (left panel) and $k = 6$ (right panel). AF function names are on the right of each boxplot experiment. The remaining part of the legend is as in Figure 2 as in the Main Text.

7

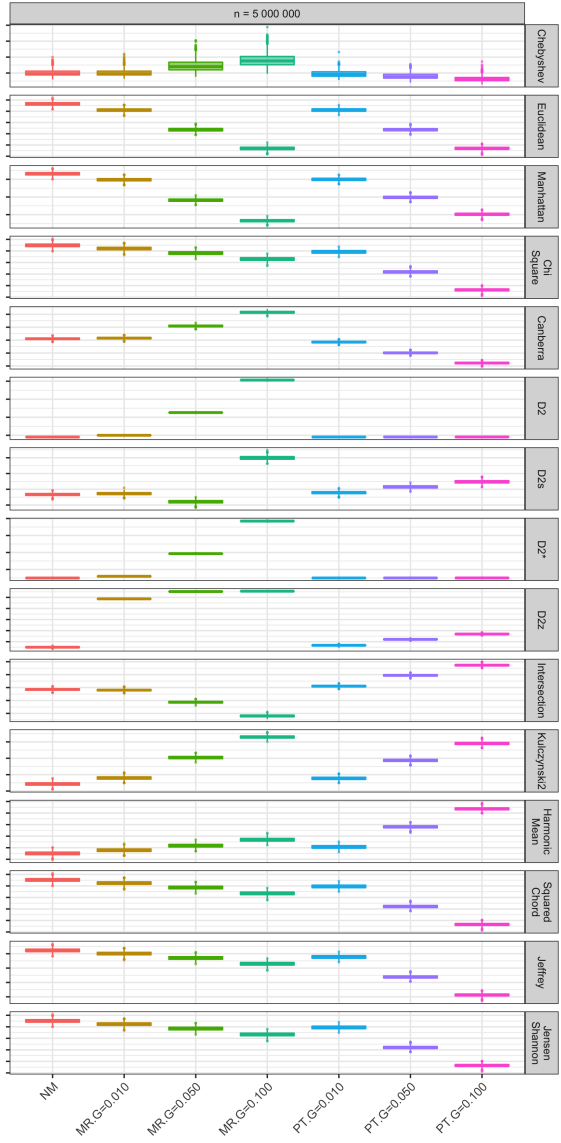
(a) k=8

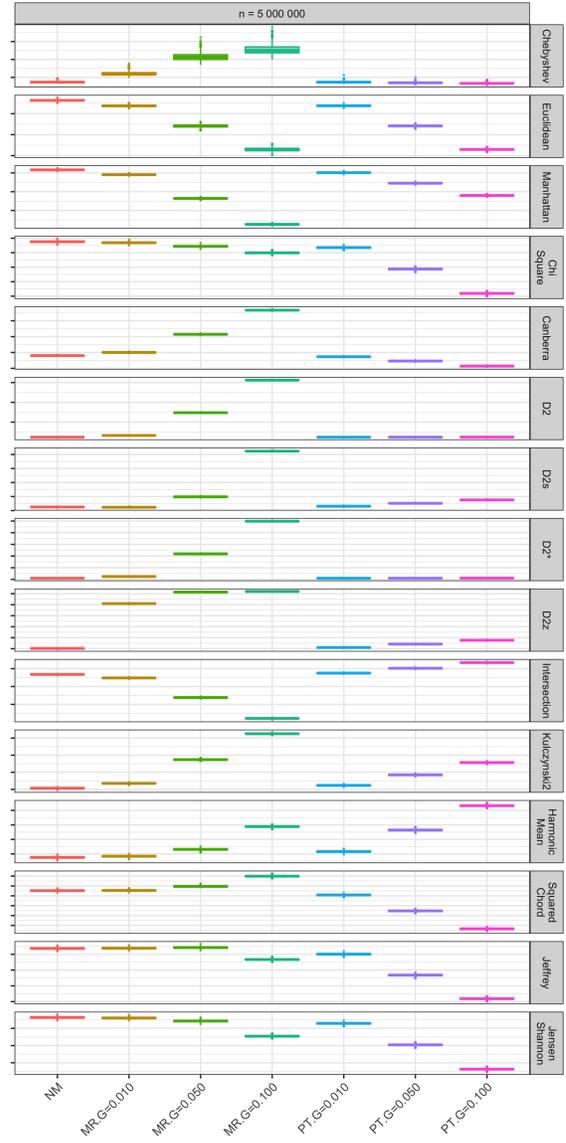
(b) k=10

Figure 5: Separation, graphical view, of the AF function values for all measures and for $n = 5\,000\,000$ and $k = 8$ (left panel) and $k = 10$ (right panel). AF function names are on the right of each boxplot experiment. The remaining part of the legend is as in Figure 2 as in the Main Text.

8

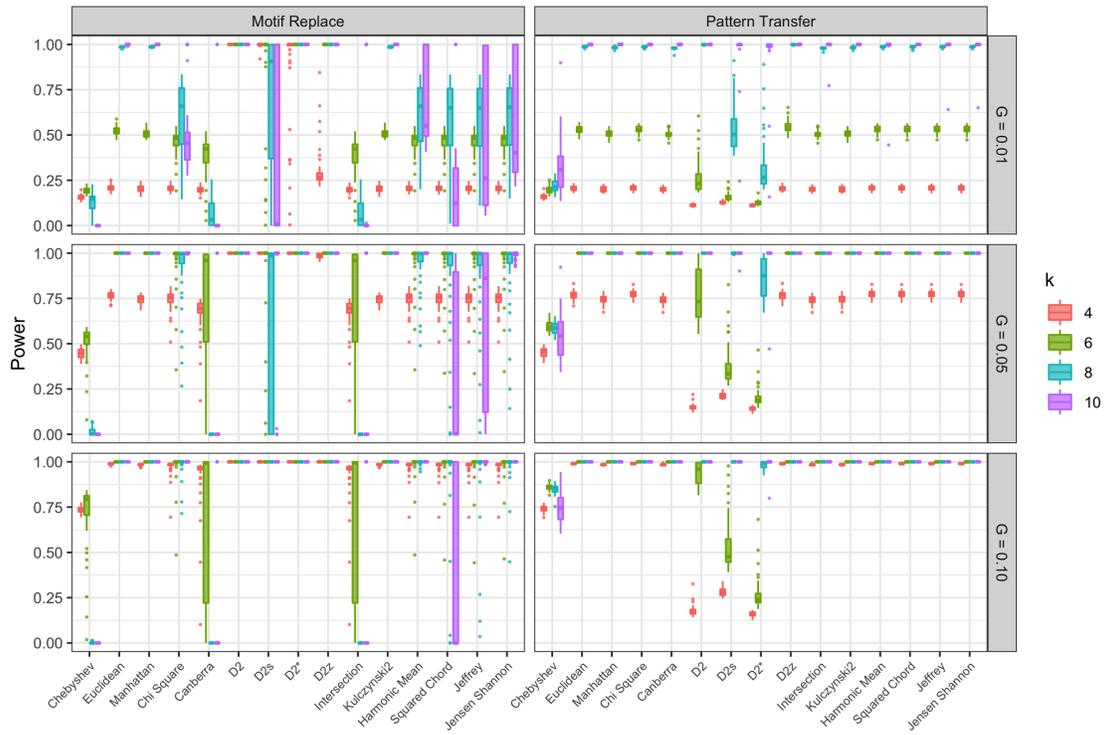

Figure 6: Distribution of the proportion of true positives cumulatively by length, for the long sequences scenario, for each AF and $\gamma = 0.01, 0.05, 0.1$, for both Alternative Models, represented by boxplots colored according to $k$.

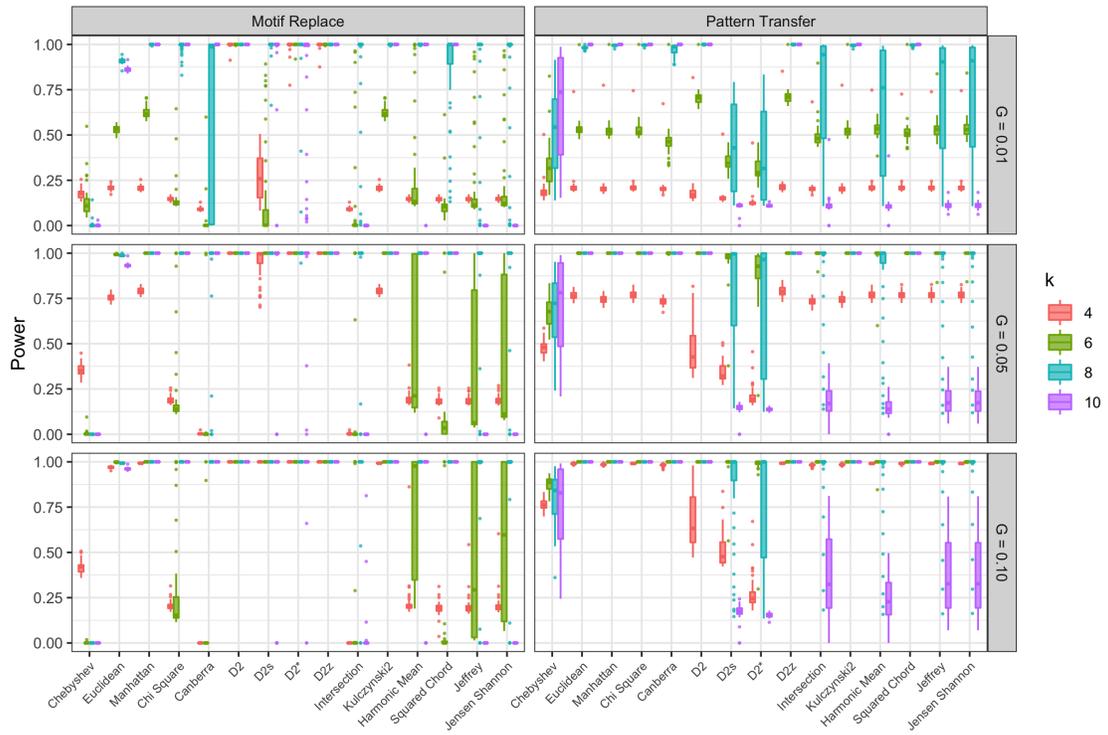

Figure 7: Distribution of the proportion of true positives cumulatively by length, for the short sequences scenario, for each AF and $\gamma = 0.01, 0.05, 0.1$, for both Alternative Models, represented by boxplots colored according to $k$.

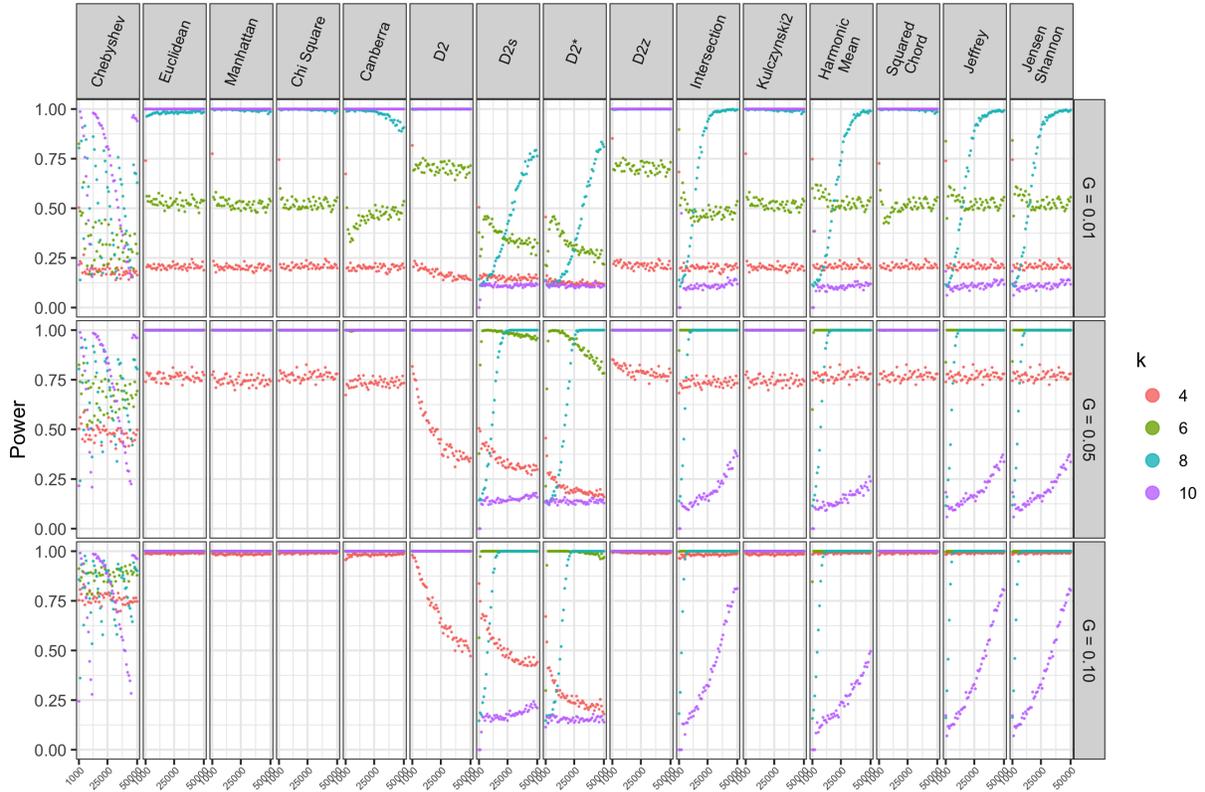

(a) Pattern Transfer

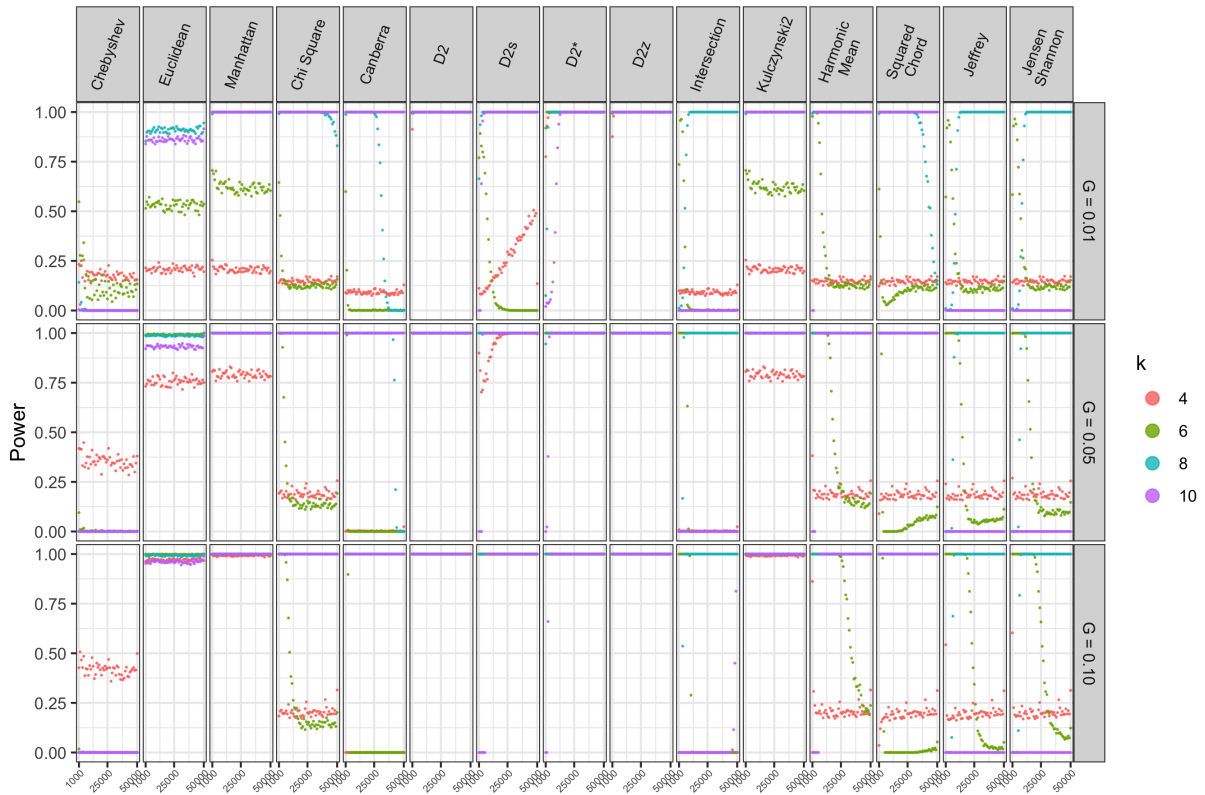

(b) Motif Replace

Figure 8: Power Analyses for the short sequences scenario. Upper Panel: Power trend for the $PT$ Alternative model. Lower Panel: Power trend for the $MR$ Alternative model. For each AF and $\gamma = 0.01, 0.05, 0.1$ the power levels obtained across different values of $n$ is reported and colored according to the value of $k$.



% \begin{thebibliography}{}

% \end{thebibliography}
\end{document}